\documentclass[12pt, a4paper]{article}

\usepackage{amsmath}
\usepackage{amsfonts}
\usepackage{amssymb}
\usepackage{amsthm}
\usepackage{pdfpages}
\usepackage{listings}
\usepackage{pgfplots}
\usepackage{a4wide}
\usepackage{comment}
\usepackage{multirow}

\newcommand{\eqn}{\begin{eqnarray}}
\newcommand{\eqnd}{\end{eqnarray}}

\newcommand{\dX}{\,\text{d$X$}}

\newcommand{\ds}{\,\text{d$s$}}
\newcommand{\dt}{\,\text{d$t$}}

\newcommand{\dW}{\,\text{d$W$}}

\newcommand{\dL}{\,\text{d$L$}}

\newcommand{\argmin}[1]{\rm{argmin}}

\def\harf{\hbox{$\frac{1}{2}$}}

\begin{document}
	
\title{
Numerical Valuation of Derivatives in\\ High-Dimensional Settings via PDE Expansions}
\author{Christoph Reisinger\footnote{Mathematical Institute and Oxford-Man Institute for Quantitative Finance, University of Oxford, 24--29 St Giles, Oxford, OX1 3LB, United Kingdom, \{reisinge,wissmann\}@maths.ox.ac.uk} \, and Rasmus Wissmann\footnotemark[\value{footnote}] \footnote{Research supported by EPSRC and Nomura via a CASE Award, the German National Academic Foundation, and St. Catherine's College, Oxford.}}
\date{\today}
\maketitle

\begin{abstract}
In this article, we propose a new numerical approach to high-dimensional partial differential equations (PDEs) arising in the valuation of exotic derivative securities. The proposed method is extended from \cite{RW07} and uses principal component analysis (PCA) of the underlying process in combination with a Taylor expansion of the value function into solutions to low-dimensional PDEs. The approximation is related to anchored analysis of variance (ANOVA) decompositions and is expected to be accurate whenever the covariance matrix has one or few dominating eigenvalues. A main purpose of the present article is to give a careful analysis of the numerical accuracy and computational complexity compared to state-of-the-art Monte Carlo methods on the example of Bermudan swaptions and Ratchet floors, which are considered difficult benchmark problems. 
We are able to demonstrate that for problems with medium to high dimensionality and moderate time horizons the presented PDE method delivers results comparable in accuracy to the MC methods considered here in similar or (often significantly) faster runtime.
\end{abstract}

\section{Introduction}

In most common models, the values of financial derivatives are equivalently characterised as the expected value of a payoff functional under some stochastic process
or the solution of an associated partial (integro-)differential equation.
The two dominant classes of numerical methods in derivative pricing are therefore Monte Carlo methods (see, e.g., \cite{G03}) for estimating the expectation via simulation and discretisation methods (see, e.g.,
\cite{AP05, TR00}) for approximating the solution to the respective PDE (where we include lattice, spectral and Fourier methods in the latter group for the properties we shall discuss now).
Simulation methods are well suited to track path-dependent quantities which determine the payoff of exotic derivatives, and scale favourably with the dimension of the process.
However, the convergence in the number of samples is slow and they require additional approximations to early exercise strategies.
Conversely, conventional PDE discretisation methods incorporate early exercise features easily and allow fast convergence in the number of nodes used in each direction, which makes them very efficient for low-dimensional problems, but they become intractable as the dimensionality increases.

The effort to solve $N$-dimensional PDEs numerically with standard grid-based methods grows exponentially with $N$ and even more sophisticated PDE methods tailored to high-dimensional approximation, such as those based on sparse grids, are typically not able to deal with practical problems where $N$ exceeds about five to eight, see \cite{heinecke12highly-parallel,HKSW10,LO08,RW07}.
Given especially the advantages in dealing with early exercise,
it would be not only of academic interest but also practically very relevant, to be able to solve generic high-dimensional derivative pricing problems with PDE methods.

In this paper, we adapt an approach from \cite{RW07} which computes an approximate solution of an $N$-dimensional PDE by solving $O(N^p)$ PDEs of maximum dimension $d\ll N$. In fact, we will see that $p=1$ and $d=2$ is usually sufficient for practically adequate accuracy.
The underlying principle of this and related approaches is an anchored ANOVA-type decomposition (see \cite{RA99}) of a solution $u(z)$, $z\in \mathbb{R}^N$, into
\begin{eqnarray}
\nonumber
\!\!\!\!\! u(z) &\!\!=\!\!& u_0(a) + \sum_{i=1}^N u_i(a;z_i) + \!\!\!\!\sum_{\scriptsize\begin{array}{c}i,j=1 \\ i<j\end{array}}^N \!\!\!\! u_{i,j}(a;z_i,z_j) + \ldots + u_{1,\ldots,N}(a;z_1,\ldots,z_N) \\
\!\!\!\!\!\! &\!\!\!=\!\!\!& \sum_{\scriptsize \begin{array}{c} v \subseteq \{1,\ldots,N\}\end{array}} u_v(a;z^v),
\label{anova}
\end{eqnarray}
where we associate $u_{\emptyset}$ with $u_0$, $u_{\{i\}}$ with $u_i$ etc.
The terms on the right-hand side each only depend on a subset of the coordinates, $z^v=(z_{i_1},\ldots, z_{i_{|v|}})$, and a chosen
`anchor' $a=(a_1,\ldots,a_N)$.
This has been successfully applied to quadrature problems from finance in \cite{GH10}, and its relation to the PDE expansions in \cite{RW07}, which form the basis for the present work, is highlighted in \cite{R12} and \cite{SGW12}.

Key to the efficiency of this approximation as a numerical method is that the relative importance of $u_v$ decays rapidly with increasing $|v|$, as $|v|$ is the dimension of the coordinate space of $u_v$.
This can be achieved by a coordinate transformation of the underlying stochastic process and of the corresponding forward or backward PDE.
Optimal linear transformations taking into account the payoff function are analysed in \cite{IT06}, while here we consider
the principal components of the covariance matrix $\Sigma$ of the Brownian driver of the process.
The accuracy of the approximate solution
obtained by truncating (\ref{anova}) after a small number of terms with small $|v|$
then depends largely on the (relative) sizes of the eigenvalues $\lambda_i$ of $\Sigma$, $1\leq i\leq N$.
This will be motivated in Section \ref{sec:PCAApproach} by expanding the value function in  $\lambda_i$.
We follow here \cite{RW07}, who first introduced this idea for vanilla basket options.

In this article, we demonstrate the wider applicability in situations where no closed-form solution is known and accurate Monte Carlo estimates are difficult to obtain.
A prime candidate for using this technique in practice is the LIBOR market model for the joint evolution of LIBOR rates with different tenors.
To value path-dependent products such as TARNs (Targeted Accrual Redemption Notes), Snowballs or Ratchets, and early exercise options such as Bermudan swaptions,
indeed the whole yield curve has to be taken into consideration, which makes the problem genuinely high-dimensional for long enough maturities.
The PCA-ANOVA-based PDE approach presented here is very well suited to this setting even in high dimensions, because 
LIBORs with similar tenors are strongly correlated, such that one observes a fast decay of the eigenvalues, as is seen from Fig.\ \ref{fig:lambda11} in Section \ref{sec:PCALMM}.
On the example of Bermudan swaptions,
even when including the first order terms with $|v|\le 1$ alone, only a mild loss of accuracy is observed as the dimensionality, determined by the number of LIBORs considered, ranges up to 50--60. 
This deterioration appears to be an effect mostly of the time to maturity rather than the dimension increase alone. For longer running contracts, the higher order terms in (\ref{anova}) become more relevant.

A similar decay of accuracy for longer maturities is observed with the commonly used Monte Carlo method presented in \cite{A04}. There, the necessary restriction of the class of exercise strategies there produces 
a gap between lower and (dual) upper bounds which widens as the maturity increases.
The accuracy of these Monte Carlo results is comparable with the PDE ones, which are obtained in a small fraction of the computational time.
Additionally, the expansion (\ref{anova}) implicitly defines a systematic accuracy improvement and is relatively straightforward to implement. We study this in Section \ref{subsec:higher-order}.

Overall, in this paper we
\begin{itemize}
\item
extend the PDE expansion method for derivative pricing from simple, log-normal equity basket models to complex, practically relevant applications with high-dimensional underlying processes, in particular path-dependent and early-exercise options on the LIBOR curve;
\item
benchmark the PDE expansion method against widely used Monte Carlo methods for options on the yield curve and thereby demonstrate for the first time that the PDE expansion method can outperform state-of-the-art Monte Carlo methods for such complex and high-dimensional applications;
\item
present a systematic and generic approach to construct higher order approximations and give numerical results demonstrating clearly the accuracy improvement achieved.
\end{itemize}

The rest of this paper is organised as follows: Section \ref{sec:PCAApproach} introduces the PCA-based PDE expansion method, and Section \ref{sec:anova} discusses its relation to anchored ANOVA decompositions. In Section \ref{sec:PCALMM} we apply the approach to the LIBOR Market Model, and in Section \ref{sec:Num} show numerical results for two LIBOR derivatives, Bermudan swaptions and Ratchet floors. Section \ref{sec:Con} summarizes the results and discusses extensions.

\section{A PCA-based PDE expansion method}\label{sec:PCAApproach}

\subsection{Basic PDE formulation and PCA}


Consider asset value processes $X_i$, $1\leq i\leq N$, satisfying
\begin{equation}\label{eq:generaldynamics1}
	\dX_i / X_i =  \mu_i(X,t) \dt + \sigma_i(X,t) \dW_i^\mathcal{Q}
\end{equation}
on a probability space $\{\Omega, \mathcal{F}, \mathcal{P}\}$ with filtration $\{\mathcal{F}_t\}, t\in[0,T]$, $T\in\mathbb{R}^+\cup\infty$.
Here $\sigma: \mathbb{R}^N \times [0,T] \rightarrow \mathbb{R}^{N,+}_0$ is the volatility, $\mu: \mathbb{R}^N \times [0,T] \rightarrow \mathbb{R}^N$ is the drift, $W^\mathcal{Q}$ is a standard Brownian motion under the risk-neutral measure $\mathcal{Q}$ and $\rho:  \mathbb{R}^N \times [0,T] \rightarrow \mathbb{R}^{N\times N}$ is the correlation matrix, i.e.,
\begin{equation}\label{eq:generaldynamics2}
	\langle \dW_i^\mathcal{Q}, \dW_j^\mathcal{Q} \rangle = \rho_{ij} \dt \;\;\; \forall i,j\in {1,\ldots,N} .
\end{equation}
A European option is characterised by its payout function $G: \mathbb{R}^N \rightarrow \mathbb{R}$, which determines the amount $G(X_T)$ its holder receives at time $t=T$. The arbitrage-free value of the option relative to the num\'eraire $\mathcal{N}$ is then
\begin{equation}
V(t, X(t)) 
= \mathbb{E}^{\mathcal{Q}}\left(\left.\frac{G(X(T))}{\mathcal{N}(T)}\right| \mathcal{F}_t\right),
\end{equation}
assuming that standard technical conditions hold\footnote{See, e.g., \cite{F07}.}. Here $G(\cdot)$ is the absolute payoff at time $T$. By the Feynman-Kac theorem, $V$ satisfies the parabolic PDE
\begin{equation}\label{eq:Feynman-Kac-PDE}
	\frac{\partial V}{\partial t} + \sum_{i=1}^N\mu_i x_i\frac{\partial V}{\partial x_i} + \sum_{i,j=1}^N\harf \sigma_i\sigma_j\rho_{ij} x_ix_j\frac{\partial^2 V}{\partial x_i \partial x_j} = 0
\end{equation}
on $\mathbb{R}^N \times[0,T]$ with final condition
\begin{equation}\label{eq:Feynman-Kac-PDEBoundary}
	V(x,T) = g(x) \;\;\; \forall x\in\mathbb{R}^N ,
\end{equation}
where for simplicity of notation we have used the relative payoff $g(\cdot) = G(\cdot)/\mathcal{N}(T)$. Equation (\ref{eq:Feynman-Kac-PDEBoundary}) naturally generalises to the Bermudan and Ratchet cases discussed later, which are modelled by the introduction of additional, intermediate conditions at a fixed, finite set of tenor times $T_1, \ldots, T_N$.

Assume now that $\rho$ and $\sigma$ are constant and $\mu$ a function of $t$ alone.
Let $\Sigma$ be the covariance matrix, $\Sigma_{ij} = \sigma_i\rho_{ij}\sigma_j$ for all $1\leq i,j\leq N$. Let $Q\in\mathbb{R}^{N\times N}$ be the orthogonal matrix of eigenvectors of $\Sigma$ and let the eigenvalues $\lambda_i$ be sorted in descending order, i.e., $\lambda_1 \geq \lambda_2 \geq \ldots \geq \lambda_N \geq 0$. 
Then the coordinate transformation
\begin{equation}\label{eq:trafo1}
	\tau = T - t \;\;,\;\; z = Q^T \ln(x) + \beta(\tau),
\end{equation}
where
\begin{equation}\label{eq:trafo2}
	\beta_i(\tau) = - \sum_{j=1}^N Q_{ji} \left(\frac{\tau\sigma^2_j}{2} + \int_0^\tau \mu_j(s) \ds \right), \;\;\; 1\leq i\leq N,
\end{equation}
leads to
\begin{equation}\label{eq:PDEHeat}
\frac{\partial u}{\partial \tau } - \frac{1}{2}\sum_{i=1}^N \lambda_i \frac{\partial^2 u}{\partial z^2_i} = 0 \;\;\; \forall (\tau,z)\in [0,T]\times \mathbb{R}^N,
\end{equation}
where $V(t,X)=u(\tau,z)$ and
\begin{equation}\label{eq:PDEHeatBoundary}
u(z,0) = g\left(\exp[Q z]\right) .
\end{equation}
Here the rotation with $Q^T$ eliminates mixed derivatives and the translation by $\beta(\tau)$ eliminates the first order terms. This can be seen by a straightforward calculation of the partial derivatives in the new coordinates (see also \cite{R04}).

\subsection{Taylor expansion}
\label{subsec:taylor}

Consider now $u$ as a function also of the vector $\lambda=(\lambda_1,\ldots,\lambda_N)$ of eigenvalues.
For any point $\lambda^0\in\mathbb{R}^N$ and $s>1$, we can define $\delta\lambda \equiv \lambda-\lambda^0$ and can formally write down the $s$-th order Taylor expansion at $\lambda^0$ as
\begin{eqnarray}
u(z,\tau;\lambda) &=& u(z,\tau;\lambda^0) + \sum_{i_1=1}^N \delta\lambda_{i_1} \frac{\partial u}{\partial \lambda_{i_1}}(z,\tau;\lambda^0) \nonumber + \ldots \\
&& \hspace{-0.5cm}  + \sum_{i_1,\ldots,i_s=1}^N \!\!\!\!\! \frac{\delta\lambda_{i_1}\cdot \ldots \cdot \delta\lambda_{i_s}}{s!}\frac{\partial^s u}{\partial \lambda_{i_1}\ldots \partial \lambda_{i_s}}(z,\tau;\lambda^0) + O\left(\|\delta\lambda\|^{s+1}\right) .
\label{taylor}
\end{eqnarray}
The error term is justified for sufficient regularity of $u$. A typical choice of expansion point would be $\lambda^0 = (\lambda_1,\ldots,\lambda_r,0,\ldots,0)$ for some $r\geq 1$. 

We can then choose suitable finite difference approximations $\Delta^{(i_1,\ldots,i_m)}(u)$ to each partial derivative
with respect to $\lambda_{i_1},\ldots,\lambda_{i_m}$. For example, Hilber et al.\ \cite{HKSW10} propose to use high order compact finite difference stencils introduced in \cite{L92}, while we use stencils based on Lagrangian interpolation as given in Table \ref{tab:stencils}.

Choosing $\delta\lambda_i$ as stepsize in direction $i$ and denoting for each $m$ by $t_m+1$ the lowest approximation order of any $\Delta^{(i_1,\ldots,i_m)}$, we have
\begin{equation}
\sum_{i_1,\ldots,i_m=1}^N \!\!\!\! \frac{\delta\lambda_{i_1}\cdot \ldots \cdot \delta\lambda_{i_m}}{m!}\frac{\partial^m u}{\partial \lambda_{i_1}\ldots \partial \lambda_{i_m}}(z,\tau; \lambda^0) = \!\!\!\! \sum_{i_1,\ldots,i_m=1}^N \!\!\!\! \Delta^{(i_1,\ldots,i_m)}(u;z,\tau;\lambda,\lambda^0) + O\left(\|\delta\lambda\|^{t_m+1}\right),
\label{fd}
\end{equation}
making explicit all arguments the finite difference approximation depends on.
We additionally set $\Delta^{0}(u;z,\tau;\lambda,\lambda^0) \equiv u(z,\tau;\lambda^0)$. 

The finite difference approximation will contain the values $u(z,\tau;\lambda')$ for different values of $\lambda'$, which depend on $\lambda$, $\lambda^0$, and 
the finite difference formula itself. 
For all sensible finite difference approximations to derivatives of mixed order $m$, the number of non-zero elements of $\lambda'$ will be $m\ll N$ plus the number of non-zeros of $\lambda^0$.
The computation of $u(z,\tau;\lambda')$ for a $\lambda'$ with $k$ non-zero components can be accomplished by the solution of a $k$-dimensional PDE of the form
\begin{equation}
\label{lambdadash}
\frac{\partial u}{\partial \tau } - \frac{1}{2}\sum_{i=1}^N \lambda'_i \frac{\partial^2 u}{\partial z^2_i} \; = \; \frac{\partial u}{\partial \tau } - \frac{1}{2}\sum_{i=1,\lambda'_i\ne 0}^N \lambda'_i \frac{\partial^2 u}{\partial z^2_i} \;=\;  0
\end{equation}
instead of the full $N$-dimensional one.
Insertion of (\ref{fd}) in (\ref{taylor}) gives us
\begin{eqnarray}\label{eq:Extension1}
u(z,\tau;\lambda) &=& \Delta^{0}(u;z,\tau;\lambda,\lambda^0) + \sum_{m=1}^s\sum_{i_1,\ldots,i_m=1}^N \Delta^{(i_1,\ldots,i_m)}(u;z,\tau;\lambda,\lambda^0) \nonumber\\
&& \hspace{5 cm} + \sum_{m=1}^s O\left(\|\delta\lambda\|^{t_m+1}\right) + O\left(\|\delta\lambda\|^{s+1}\right).
\end{eqnarray}
The overall approximation order is $t+1$, where $t = \min\{t_1,\ldots,t_m,s\}$, and the error is $\epsilon = O\left(\|\delta\lambda\|^{t+1}\right)$.

\subsection{First-order, first eigenvalue case}\label{subsec:1o1d}

A good choice of $\lambda^0$ and the number of terms to include in the Taylor expansion depends on the problem at hand.
However, it is a common feature of processes with strong correlation that there is a dominant eigenvalue which is much larger than the rest of the spectrum.
This is also the case for the model parameters illustrated in Fig.~\ref{fig:lambda11} in Section \ref{sec:PCALMM}.

This motivates to expand up to first order, $n=1$, around $\lambda^0 = (\lambda_1,0,\ldots,0)$.
Using a simple first-order forward finite difference approximation
\begin{equation}
	\Delta^{(i)}(u;z,\tau;\lambda,\lambda^0) = \delta\lambda_{i}\,  \frac{u(z,\tau;\lambda^0+\delta \lambda_i e_i)-u(z,\tau;\lambda^0)}{\delta\lambda_{i}}
\end{equation}
to the first derivative, where $e_i$ is the $i$-th canonical basis vector, $i>1$, we get a scheme with overall order $t = \min\{2,1 + 1\} = 2$. The corresponding error is of size $O(\|\delta\lambda\|^2) = O(\lambda_2^2 + \ldots + \lambda_N^2)$. To evaluate (\ref{eq:Extension1}) up to $n=1$, we have to solve the one-dimensional PDE 
\begin{equation}
\label{onedpde}
\frac{\partial u}{\partial \tau } - \frac{1}{2} \lambda_1 \frac{\partial^2 u}{\partial z^2_1} = 0
\end{equation}
and the $N-1$ two-dimensional PDEs 
\begin{equation}
\label{twodpde}
\frac{\partial u}{\partial \tau } - \frac{1}{2} \lambda_1 \frac{\partial^2 u}{\partial z^2_1} - \frac{1}{2} \lambda_i \frac{\partial^2 u}{\partial z^2_i} = 0 ,
\end{equation}
$2\leq i\leq N$, and obtain the approximate solution
\begin{eqnarray}
\label{case11}
	u^{(1,1)}(z,\tau;\lambda) &=& 
	u(z,\tau;\lambda^0) + \sum_{i=2}^N \left(u(z,\tau;\lambda^0+\lambda_i e_i) - u(z,\tau;\lambda^0)\right) \\
	&=& (2-N) \, u(z,\tau;\lambda^0) + \sum_{i=2}^N u(z,\tau;\lambda^0+\lambda_i e_i).
\end{eqnarray}
The superscript $(1,1)$ of $u$ in (\ref{case11}) indicates that $\lambda^0$ has one non-zero element, and we are truncating the Taylor expansion after the first term.
The largest dimension of any PDE to be solved is 1+1=2.

\section{Generalisations and relation to anchored ANOVA}
\label{sec:anova}

Anchored ANOVA-decompositions are used in \cite{GH10} to obtain dimension-adaptive approximations to option values expressed as integrals over high-dimensional spaces;
\cite{SGW12} point out a relation to the expansions from \cite{RW07} by utilising the integral representation of the solution to the heat equation; \cite{R12} discusses these ideas
jointly in the PDE context.

We formulate the problem in slightly more general terms here as befits the applications later on.
Consider the situation where $Z=(Z_1(t),\ldots, Z_N(t))_{t\ge 0}$ is an $N$-dimensional Markov process and where the time $t$ value $u$ of a contingent claim is fully determined by $Z(t)$ and
the time to maturity $T-t$. We therefore write this value
as $u(Z(t),T-t)$. 
Define, for a set $v=\{i_1,\ldots,i_m\} \subseteq \{1,\ldots, N\}$, auxiliary processes $Z^v=(Z^v_1,\ldots,Z^v_N)$ which are ``frozen'' in the coordinates with indices not in $v$, that is, $Z^v_i(t)=Z_i(0)$ for all $t\ge 0$, $i\notin v$, and  we impose that the joint law of $\{Z_i^v, i\in v\}$, is identical to the joint law of $\{Z_i, i\in v\}$, \emph{given} that $Z_j(t) = Z_j(0)$ for all $j\notin v$.
To be more specific, in the common case where $Z$ is defined through
a stochastic differential equation (SDE) of the type
\begin{eqnarray}
\label{fullsde}
\text{d} Z_{i}(t) = \mu_i(Z(t),t) \, \text{d}t + \sigma_i(Z(t),t) \, \text{d}W_i(t),
\end{eqnarray}
we define $Z^v$ by
\begin{eqnarray}
\label{vfreeze}
\text{d} Z^v_{i}(t) = \left\{
\begin{array}{rl}
\mu_{i}(Z^v(t),t) \, \text{d}t + \sigma_{i}(Z^v(t),t) \, \text{d}W_{i}(t) & i \in v, \\
0 & \text{else},
\end{array}
\right.
\end{eqnarray}
which are constant in directions $i\in \{1,\ldots,N\}\backslash v$, i.e., $Z^v_{i}(t)= Z^v_{i}(0)$ for those $i$.

A particular example studied in \cite{GH10,SGW12}, which is related to the valuation of European-style derivatives, is
\begin{eqnarray}
\label{fullexp}
u(z,\tau) = u(z,T-t)=\mathbb{E}[g(Z(T))|Z(t)=z], 
\end{eqnarray}
where $\tau=T-t$ is the time to maturity, $g$ is the payoff function and the expectation is taken with respect to an explicitly known probability measure.
In that case, we define approximations based on the process $Z^v$ in (\ref{vfreeze}) as
\begin{eqnarray}
\label{vexp}
\widehat{u}_v(a;z^v,\tau) = \mathbb{E}[g(Z^v_T)| Z^v_i(t)=z_i \; \forall \,  i\in v; Z^v_i(t)=a_i \; \forall \, i\notin v],
\end{eqnarray}
where $a \in \mathbb{R}^N$ with $a_i = Z_{i}(0)$, $z^v=(z_{i_1},\ldots, z_{i_{|v|}})$,
i.e., we anchor the solution at the initial value of this stochastic process.

The forward and backward PDEs for processes of the type (\ref{fullsde}) are second order linear parabolic. To get from
the PDE for (\ref{fullexp}) to the one for (\ref{vexp}), the coefficients of all derivatives in directions $x_i$ are set to zero for $i\notin v$, as per (\ref{vfreeze}).

The coordinates $z$ in (\ref{eq:trafo1}) were chosen specifically as the principal components of the covariance matrix of a diffusion process $X$, for constant $\mu$ and $\sigma$, in which case
$\mu_{Z,i}=0$ and $\sigma_{Z,i}^2 = \lambda_i$.
A similar construction is used and analysed in \cite{HKSW10}.
The PDE satisfied by $\widehat{u}_v$ in (\ref{vexp}) is (\ref{lambdadash}) with $\lambda'$ set to 
\begin{equation}
\label{lambdav}
\lambda^v = \sum_{i\in v} \lambda_i e_i.
\end{equation}

As the solution $\widehat{u}_v$, $v=\{i_1,\ldots,i_{|v|}\}$, only depends on the sub-vector of coordinates $z^v=(z_{i_1},\ldots, z_{i_{|v|}})$ 
non-trivially,
\begin{eqnarray}
\label{surplus}
u_v(a;z^v,\tau) &=& \widehat{u}_v - \sum_{w\subset v} u_w \\
&=&  \sum_{w\subseteq v} (-1)^{|w|-|v|} \widehat{u}_w
\end{eqnarray}
is well-defined and gives a suitable anchored ANOVA decomposition of $u$ (see \cite{GH10} and the references therein) 
as given in (\ref{anova}), where $\tau$ is an additional argument of all terms.

For optimal stopping problems, such as the Bermudan swaptions studied later, the analogue to (\ref{fullexp}) and (\ref{vexp}) is
\begin{eqnarray}
u(z,\tau) &=& \sup_{\mathcal{T} \in \mathbb{T}} \mathbb{E}[g(Z(\mathcal{T}))|Z(t)=z], \\
\widehat{u}_v(a;z^v,\tau) &=& \sup_{\mathcal{T} \in \mathbb{T}^v} \mathbb{E}[g(Z^v(\mathcal{T}))|Z^v_i(t)=z_i \; \forall \,  i\in v; Z^v_i(t)=a_i \; \forall \, i\notin v],
\end{eqnarray}
where $\mathbb{T}, \mathbb{T}^v$ are suitable sets of stopping times. 

For path-dependent options, the process $Z$ has to be set up to include the path-dependent quantity in order to bring it back into the assumed Markovian framework.
In Section \ref{subsec:ratfloor},
we demonstrate this on the example of a ratchet floor. 
As the path-dependent quantity is reset at discrete time points, the corresponding component of $Z$ is a jump-process instead of following an SDE of the form
(\ref{fullsde}).

The general principle is that we define $\widehat{u}_v$ as the ``solution of the problem with $Z$ replaced by $Z^v$''.

To re-iterate, the key is that $Z^v$ changes only in $|v|$ dimensions and is constant in the remaining $N-|v|$ dimensions. This means that $u_v$ can be found by solving $|v|$-dimensional (e.g., PDE) problems instead of the $N$-dimensional one.

The link between (\ref{anova}) and (\ref{eq:Extension1}) can now be established
if we pick $\lambda^0=0$, set $v=\{i_1,\ldots,i_k\}$, $\lambda^v$ as in (\ref{lambdav}), and, inductively (skipping $z$ and $\tau$ as argument of $\Delta$ for brevity),
\begin{eqnarray}
\label{firstfd}
	\Delta^{(i_1)}(u;\lambda,\lambda^0) &=& \delta\lambda_{i} \frac{u(z,\tau;\lambda^0+\delta\lambda_{i_1} e_{i_1})-u(z,\tau;\lambda^0)}{\delta\lambda_{i_1}}, \\
\hspace{-0.6 cm} 	\Delta^{(i_1,\ldots,i_{k},i_{k+1})}(u;\lambda,\lambda^0) &=& \Delta^{(i_{k+1})}(\Delta^{(i_1,\ldots,i_{k})};\lambda,\lambda^v) \\	
	&=& \delta\lambda_{i_{k+1}}\frac{ \Delta^{(i_1,\ldots,i_{k})}(u;\lambda,\lambda^v+\delta\lambda_{i_{k+1}} e_{i_{k+1}})- \Delta^{(i_1,\ldots,i_{k})}(u;\lambda,\lambda^v)}{\delta\lambda_{i_{k+1}}} \\
	&=& \sum_{w\subseteq v\cup \{k+1\}} (-1)^{|w|-|v|-1} u(z,\tau;\lambda^w),
\end{eqnarray}
for $i_l\neq i_m$, $1\le l < m\le k+1$, and 0 otherwise. Only terms of mixed first order are present and absorb all higher order terms -- see next paragraph.
Then, the precise relation between ANOVA terms and the finite difference approximation to the Taylor expansion is
\begin{eqnarray}
\widehat{u}_v(z;z^v,0) &=& u(z,0;\lambda^v), \\
u_v(z;z^v,0) &=& \Delta^{(i_1,\ldots,i_{k})}(u;z,0;\lambda,\lambda^0).
\end{eqnarray}
The relation between ANOVA and multi-variate Taylor expansions in the coordinates is discussed in \cite{G06}.
The twist here is to apply the expansion in $\delta \lambda_i$ instead of $z_i$.

A further point to note is that if the above expansion is truncated to include terms up to $|v|=n<N$, it is only of first order accurate in $\delta \lambda_i$.
For relatively large $\delta \lambda_i$ and smooth solutions, the inclusion of higher order Taylor terms in individual and mixed directions and higher order finite difference formulae 
may be preferable as we will see in Section \ref{subsec:ratfloor}. The extra cost is small
as typically the dimensionality of PDEs involved will not increase.
What distinguishes the above expansion from other finite difference approximations is that it is an exact decomposition, i.e., if we include all terms up to degree $N$, we recover the exact solution irrespective of its smoothness.

In a variation to (\ref{anova}), we can consider a decomposition, where in addition to the anchor $a$, all contributions may
also depend on the first coordinate,
\begin{eqnarray}
\nonumber
\!\!\!\!u(z) \!&=& \!u_0^{(1)}(a;z_1) + \sum_{i=2}^N u^{(1)}_i(a;z_1;z_i) +   \!\!\!\! \sum_{\scriptsize\begin{array}{c}i,j=2\\i<j\end{array}}^N u^{(1)}_{i,j}(a;z_1;z_i,z_j) + \ldots \\
&& \hspace{4 cm} + \;\; u^{(1)}_{2,\ldots,N}(a;z_1;z_2,\ldots,z_N),
\label{anova1}
\end{eqnarray}
and, generalising this from one to $r\ge 1$ coordinates,
\begin{eqnarray}
\label{anovar}
u(z) \!\!&=&\!\!
u_0^{(r)}(a;z_1,\ldots,z_r) \;+\; 
\sum_{p=1}^{N-r} \!\!\!\! \sum_{\scriptsize\begin{array}{c}\{i_1,\ldots,i_p\}\\\subseteq\{r+1,\ldots,N\}\end{array}} u_{i_1,\ldots,i_p}^{(r)}(a;z_1,\ldots,z_r;z_{i_1},\ldots,z_{i_p}).
\end{eqnarray}
Clearly, in relation to Section \ref{subsec:taylor}, this corresponds to using $\lambda^0 = (\lambda_1,0,\ldots,0)$ and $\lambda^0 = (\lambda_1,\lambda_2,\ldots, \lambda_r, 0,\ldots,0)$, resp.,
and an adaptation of the finite difference formulae.

The goal is to find a decomposition where the contributions $u_{i_1,\ldots,i_p}^{(r)}$ decay fast with increasing $r$ and increasing $p$, in order for the approximations
\begin{eqnarray}
\label{anovars}
u^{(r,s)}(z) \!\!&=&\!\!
u_0^{(r)}(a;z_1,\ldots,z_r) \;+\; 
\sum_{p=1}^{s} \!\!\!\!  \sum_{\scriptsize\begin{array}{c}\{i_1,\ldots,i_p\}\\\subseteq\{r+1,\ldots,N\}\end{array}} u_{i_1,\ldots,i_p}^{(r)}(a;z_1,\ldots,z_r;z_{i_1},\ldots,z_{i_p})
\end{eqnarray}
for $s\le N-r$ to be accurate for small $r+s$.
The approximation from Section \ref{subsec:1o1d} corresponds to $r=s=1$.
In \cite{F10}, a natural link between ANOVA decompositions and dimension adaptive sparse grids is exploited to construct \emph{a priori}
as well as \emph{a postiori} optimal approximations to high-dimensional functions.

The effect of higher-dimensional terms in the cases $r=2$, $s=1$ and $r=1$, $s=2$ is illustrated in \cite{SGW12} for equity basket options, extending
the case $r=1$, $s=1$ in \cite{RW07}.
The data there have in common with our set-up the presence of a dominant eigenvalue, such that the case $(r,s)=(1,2)$ gives a notable improvement over $(1,1)$ for arithmetic average basket options by capturing the second order terms in the small eigenvalues, while $(r,s)=(2,1)$ does not give a big accuracy gain.

We will give numerical results up to $r=5$ and $s=3$ in Section \ref{subsec:ratfloor}, in the context of the LIBOR Market Model described in the following section.



\section{Application to the LIBOR Market Model}\label{sec:PCALMM}

We now apply the PCA-ANOVA approach to practically relevant examples from interest rate markets: LIBOR market derivatives. Forward rates will be assumed to follow the LIBOR Market Model (LMM), which is one of the most widely used models \cite{BGM97,F07,MSS97} and the basis for a variety of extensions.
The methods studied here have the potential to be applied to those as well.
Our notation and definition of the LMM follows \cite{F07}.

As traded product at time $t$ consider a (zero-coupon) bond $P(t,T)$, $T\geq0$, that pays 1 at time $T\geq t$. The forward LIBOR with fixing date $T$ and payout date $T'$ is then defined on the same probability space $(\Omega,\mathcal{F},\mathcal{P})$, as the stochastic process $L(\cdot,T,T') : [0,T] \times \Omega\rightarrow \mathbb{R}$ given by
\begin{equation}\label{eq:LIBOR}
L(t,T,T') \equiv \frac{1}{T'-T}\frac{P(t,T)-P(t,T')}{P(t,T')}.
\end{equation}
For a fixed tenor structure
\begin{equation}\label{eq:TenorStructure}
0 = T_1 < \ldots < T_N < T_{N+1} = T',
\end{equation}
the LMM now describes a finite number of forward rates 
\begin{equation}
L_i(t) \equiv L(t,T_i,T_{i+1})
\end{equation}
for $1\leq i \leq N,t\in\left[0,T_{i}\right]$.
Let $T_j - T_{j-1} = \alpha$ for all $2\leq j\leq N+1$. For example, two practically important values for $\alpha$ are $0.25$ and $0.5$ for 3-month and 6-month LIBOR. The full dynamics for each $L_i(t)$, $t \leq T_i$, under the equivalent martingale measure $\mathcal{Q}^{n+1}$, $1\leq n\leq N$, associated with choosing the bond $P(T_{n+1})$ as num\'eraire, are
\begin{equation}\label{eq:LIBORDynamics}
\dL_i(t) = \mu_i(t)L_i(t) \dt  + \, \sigma_i(t) L_i(t) \dW^{\mathcal{Q}^{n+1}}_i\!\! ,
\end{equation}
where
\begin{equation}\label{eq:LIBORDrift_1}
\mu_i(t) = \left\{ \begin{array}{c l}
- \sum_{j=i+1}^{n} {\frac{\alpha L_j(t)}{1+\alpha L_j(t)} \sigma_i(t) \sigma_j(t) \rho_{ij}(t)}  & i < n \\
0 & i=n\\
\sum_{j=i+1}^{N} {\frac{\alpha L_j(t)}{1+\alpha L_j(t)} \sigma_i(t) \sigma_j(t) \rho_{ij}(t)} & i > n
\end{array}\right. 
\end{equation}
for $t\leq T_{n+1}$ and similarly
\begin{equation}\label{eq:LIBORDrift_2}
\mu_i(t) = \sum_{j=\max\{k : T_k\leq t\}}^{N} {\frac{\alpha L_j(t)}{1+\alpha L_j(t)} \sigma_i(t) \sigma_j(t) \rho_{ij}(t)}
\end{equation}
for $t>T_{n+1}$.

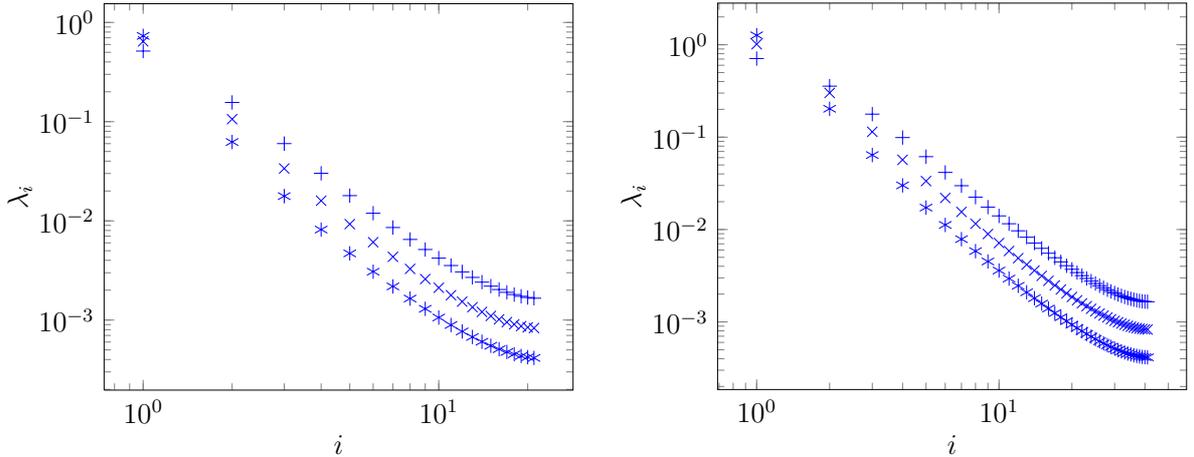
\begin{figure}[tbp]
	\begin{center}
		\begin{minipage}[l]{0.49\textwidth}
			\begin{tikzpicture}[scale=.9]
				\begin{axis}[xlabel=$i$, ylabel=$\lambda_i$, ymode=log, xmode=log]
					\addplot+[only marks,mark=asterisk,mark options={color=blue,fill=blue},mark size=3] coordinates { 
					(	1	,0.731592742543562	)
					(	2	,0.0623265458297981	)
					(	3	,0.0177497444857903	)
					(	4	,0.00817052180913912	)
					(	5	,0.00469982759669041	)
					(	6	,0.00307243652306172	)
					(	7	,0.00218351112499891	)
					(	8	,0.00164656571414803	)
					(	9	,0.00129839973630895	)
					(	10	,0.00106055483233780	)
					(	11	,0.000891553295878553	)
					(	12	,0.000767817991877791	)
					(	13	,0.000675161497015311	)
					(	14	,0.000604649503878644	)
					(	15	,0.000550447291656519	)
					(	16	,0.000508633991870642	)
					(	17	,0.000476518984978470	)
					(	18	,0.000452232798574565	)
					(	19	,0.000434475179955127	)
					(	20	,0.000422357179151031	)
					(	21	,0.000415302089327831	)

					};
					\addplot+[only marks,mark=x,mark options={color=blue,fill=blue},mark size=3] coordinates { 
					(	1	,0.644367919518585	)
					(	2	,0.105883465210391	)
					(	3	,0.0337083743778176	)
					(	4	,0.0159546035187038	)
					(	5	,0.00927432916317377	)
					(	6	,0.00609339811386969	)
					(	7	,0.00434240764917207	)
					(	8	,0.00328005905909019	)
					(	9	,0.00258931019249374	)
					(	10	,0.00211656979416585	)
					(	11	,0.00178023432976806	)
					(	12	,0.00153375855167813	)
					(	13	,0.00134906511634276	)
					(	14	,0.00120844055213290	)
					(	15	,0.00110030077728505	)
					(	16	,0.00101685314357116	)
					(	17	,0.000952745677736394	)
					(	18	,0.000904257373606433	)
					(	19	,0.000868798913841935	)
					(	20	,0.000844599382012230	)
					(	21	,0.000830509584561763	)
			
					};
					\addplot+[only marks,mark=+,mark options={color=blue,fill=blue},mark size=3] coordinates { 
					(	1	,0.514547308804207	)
					(	2	,0.156205198650511	)
					(	3	,0.0600582012365139	)
					(	4	,0.0301544359807705	)
					(	5	,0.0179552482702004	)
					(	6	,0.0119375567415143	)
					(	7	,0.00856387627044191	)
					(	8	,0.00649515047543545	)
					(	9	,0.00514101057050167	)
					(	10	,0.00421009770194858	)
					(	11	,0.00354571794141791	)
					(	12	,0.00305774134785060	)
					(	13	,0.00269146936276593	)
					(	14	,0.00241223857021535	)
					(	15	,0.00219730283452182	)
					(	16	,0.00203132052555230	)
					(	17	,0.00190373347775655	)
					(	18	,0.00180718931629837	)
					(	19	,0.00173656546702068	)
					(	20	,0.00168835517254770	)
					(	21	,0.00166028128200903	)

					};
				\end{axis}
			\end{tikzpicture}
		\end{minipage}
		\begin{minipage}[r]{0.49\textwidth}
			\begin{tikzpicture}[scale=.9]
				\begin{axis}[xlabel=$i$, ylabel=$\lambda_i$, ymode=log, xmode=log]
					\addplot+[only marks,mark=asterisk,mark options={color=blue,fill=blue},mark size=3] coordinates { 
					(	1	,1.26476716589185	)
					(	2	,0.202873526638480	)
					(	3	,0.0639920126003481	)
					(	4	,0.0300385893645525	)
					(	5	,0.0172838153048376	)
					(	6	,0.0112118832185954	)
					(	7	,0.00786726291294232	)
					(	8	,0.00583466151172220	)
					(	9	,0.00450911240966533	)
					(	10	,0.00359758498554587	)
					(	11	,0.00294435286640445	)
					(	12	,0.00246051999123897	)
					(	13	,0.00209236883816743	)
					(	14	,0.00180589928643398	)
					(	15	,0.00157874667459399	)
					(	16	,0.00139571434266805	)
					(	17	,0.00124618555920114	)
					(	18	,0.00112256309747903	)
					(	19	,0.00101929543977850	)
					(	20	,0.000932250557804499	)
					(	21	,0.000858302443316415	)
					(	22	,0.000795051653444815	)
					(	23	,0.000740632454518888	)
					(	24	,0.000693577216493184	)
					(	25	,0.000652719441913603	)
					(	26	,0.000617123356669045	)
					(	27	,0.000586032074715643	)
					(	28	,0.000558828954500154	)
					(	29	,0.000535008459593984	)
					(	30	,0.000514153958428713	)
					(	31	,0.000495920653715695	)
					(	32	,0.000480022348812610	)
					(	33	,0.000466221116636998	)
					(	34	,0.000454319188633030	)
					(	35	,0.000444152560642922	)
					(	36	,0.000435585941844142	)
					(	37	,0.000428508767287296	)
					(	38	,0.000422832064354071	)
					(	39	,0.000418486015778461	)
					(	40	,0.000415418101751876	)
					(	41	,0.000413591734641114	)
			
					};
					\addplot+[only marks,mark=x,mark options={color=blue,fill=blue},mark size=3] coordinates { 
					(	1	,1.01356428282483	)
					(	2	,0.300886373099096	)
					(	3	,0.114254968558891	)
					(	4	,0.0568047992677633	)
					(	5	,0.0334564078335316	)
					(	6	,0.0219536732582119	)
					(	7	,0.0155041597251797	)
					(	8	,0.0115440045883956	)
					(	9	,0.00894454424237805	)
					(	10	,0.00714917668166306	)
					(	11	,0.00585860058317185	)
					(	12	,0.00490056229579906	)
					(	13	,0.00417035989786617	)
					(	14	,0.00360143209410386	)
					(	15	,0.00314984899482557	)
					(	16	,0.00278568203532970	)
					(	17	,0.00248797862946708	)
					(	18	,0.00224171986510929	)
					(	19	,0.00203591567822766	)
					(	20	,0.00186237661680329	)
					(	21	,0.00171490107305526	)
					(	22	,0.00158872492473783	)
					(	23	,0.00148014112259263	)
					(	24	,0.00138623184419052	)
					(	25	,0.00130467673535612	)
					(	26	,0.00123361353672259	)
					(	27	,0.00117153538685102	)
					(	28	,0.00111721420197475	)
					(	29	,0.00106964286096121	)
					(	30	,0.00102799113165193	)
					(	31	,0.000991571763043206	)
					(	32	,0.000959814186513999	)
					(	33	,0.000932243976595422	)
					(	34	,0.000908466719427050	)
					(	35	,0.000888155291680038	)
					(	36	,0.000871039808583361	)
					(	37	,0.000856899686581552	)
					(	38	,0.000845557404418183	)
					(	39	,0.000836873650181218	)
					(	40	,0.000830743620956803	)
					(	41	,0.000827094303279909	)
			
					};
					\addplot+[only marks,mark=+,mark options={color=blue,fill=blue},mark size=3] coordinates { 
					(	1	,0.709062362078800	)
					(	2	,0.356499465340740	)
					(	3	,0.177455327681295	)
					(	4	,0.0989593617133644	)
					(	5	,0.0614962392060982	)
					(	6	,0.0415108428807061	)
					(	7	,0.0298024450059657	)
					(	8	,0.0224206944099896	)
					(	9	,0.0174920182396777	)
					(	10	,0.0140482575860539	)
					(	11	,0.0115523220626263	)
					(	12	,0.00968829062794841	)
					(	13	,0.00826106748019656	)
					(	14	,0.00714514406850515	)
					(	15	,0.00625692612440429	)
					(	16	,0.00553905285212667	)
					(	17	,0.00495113597296448	)
					(	18	,0.00446408999455443	)
					(	19	,0.00405655004006223	)
					(	20	,0.00371254486612629	)
					(	21	,0.00341994751457333	)
					(	22	,0.00316942075720793	)
					(	23	,0.00295368498537891	)
					(	24	,0.00276700080663581	)
					(	25	,0.00260479742794461	)
					(	26	,0.00246340180403972	)
					(	27	,0.00233983857530256	)
					(	28	,0.00223168048648124	)
					(	29	,0.00213693530558827	)
					(	30	,0.00205395947650695	)
					(	31	,0.00198139159030248	)
					(	32	,0.00191810071837564	)
					(	33	,0.00186314601401063	)
					(	34	,0.00181574495066082	)
					(	35	,0.00177524825231145	)
					(	36	,0.00174112006796430	)
					(	37	,0.00171292230580912	)
					(	38	,0.00169030231207968	)
					(	39	,0.00167298328208457	)
					(	40	,0.00166075694555921	)
					(	41	,0.00165347818897779	)
			
					};
				\end{axis}
			\end{tikzpicture}
		\end{minipage}
	\end{center}
	\caption{Eigenvalues $\lambda_i$, $1\le i\le N$, of $\Sigma$ for $N=21$ (left) and $N=41$ (right), constant volatility $c=0.2$, and $\phi=0.02065$ ($\ast$), $\phi=0.0413$ ($\times$) and $\phi=0.0826$ ($+$). For each value of $N$ and $\phi$ the eigenvalues approximately lie on a line with slope $-2$.}
	\label{fig:lambda11}
\end{figure}

Our model for the correlation structure is taken from \cite{KS06} with 
\begin{eqnarray}
\label{rho}
\rho_{ij} = \exp(-\phi|i-j|)
\end{eqnarray}
for $1\leq i,j \leq N$, $d=N-1$, $\phi > 0$ and a constant volatility $\sigma_i(t) = c = 0.2$ for $1\leq i\leq N, t\in[0,T_i]$.
The eigenvalues of the covariance matrix $\Sigma$ decrease rapidly. Figure \ref{fig:lambda11} demonstrates this for $N=21$ and $N=41$ and different values of $\phi$. 

The first eigenvalue $\lambda_1$ is significantly larger than the second and following eigenvalues. For example, for $\phi=0.0413$ we have $\lambda_1=0.64$, $\lambda_2=0.11$, $\lambda_3=0.03$ ($N=21$) and $\lambda_1=1.01$, $\lambda_2=0.30$, $\lambda_3=0.11$ ($N=41$). This motivates the use of the first order, first eigenvalue approach from Section \ref{subsec:1o1d}, i.e., $r=s=1$. In the case $N=41$, one might consider going to $r=2$ based on the eigenvalues alone, but we will see in the numerical tests for the Bermudan swaption that even with $r=1$ the result lies within the Monte Carlo bounds.

We now choose the terminal bond $P(T_{N+1})$ as num\'eraire and combine the LIBOR dynamics in equations (\ref{eq:LIBORDynamics})--(\ref{eq:LIBORDrift_2}) and our covariance structure with equation
(\ref{eq:Feynman-Kac-PDE})
to obtain a PDE satisfied by the value function of derivatives on the LIBOR curve. 
A complication arises in the transformation (\ref{eq:trafo1}) to the heat equation (\ref{eq:PDEHeat}), as 
the drift term $\beta_i$ in (\ref{eq:trafo2}) was assumed to depend only on $\tau$ whereas with $\mu_i$ as in equation (\ref{eq:LIBORDrift_1}) it also depends on $L_j$, $j>i$.

To make the PCA approach directly applicable, we first approximate the drift term. 
A common approach in practice is to ``freeze'' the drift at its initial value by setting
\begin{equation}
\mu_i(t) = - \sum_{j=i+1}^{N} {\frac{\alpha L_j(0)}{1+\alpha L_j(0)} \sigma_i \sigma_j \rho_{ij}}.
\end{equation}
This introduces an error in the drift that grows with $L_j(t) - L_j(0)$, and the approximation can be expected to be reasonably accurate for moderate values of $T_N$ and $\sigma$.
We will confirm this numerically by comparing the PDE results to Monte Carlo estimates with and without drift approximation.
A more accurate procedure is suggested in Section \ref{sec:Con}.

A second point of consideration is that $L_i(t)$ is only financially meaningful for $t\leq T_i$.
In order not to have to change the underlying set of arguments of the value function, and hence the PCA, at every tenor time $T_i$, we consider ``extended'' LIBORs which are also defined for $T_i < t\leq T_N$. In the case of constant $\rho$ and $\sigma$, a possible extension is obtained by demanding that $L_i(t)$ follows (\ref{eq:LIBORDynamics}) for all $0\leq t\leq T_N$.
Note that the exact option value does not depend on $L_i(t)$ for $t>T_i$ and is thus not affected by this extension.

Applying the first order, one-dimensional PCA ANOVA approach from Section \ref{subsec:1o1d} now leads to the approximate solution
\begin{eqnarray}\label{eq:LMMPCA}
u^{(1,1)}(z, \tau;\lambda) &=& u(z,\tau;\lambda^0) + \sum_{i=2}^N \left(u(z, \tau;\lambda^0+\lambda_i e_i) -  u(z,\tau;\lambda^0)\right) \\
&=& (2-N)\cdot u(z,\tau;\lambda^0) + \sum_{i=2}^N u(z, \tau;\lambda^0+\lambda_i e_i),
\label{eq:LMMPCA2}
\end{eqnarray}
where $\lambda^0 = (\lambda_1,0,\ldots,0)$,
\begin{equation}\label{eq:LMMtrafo1}
	\tau = T - t, \;\;\; z = Q^T \ln(L) + \beta(\tau)
\end{equation}
and
\begin{equation}\label{eq:LMMtrafo2}
	\beta_i(\tau) = -\, \tau \sum_{j=1}^N Q_{ji} \left(\frac{\sigma^2_j}{2} - \sum_{k=j+1}^{N} {\frac{\alpha L_k(0)}{1+\alpha L_k(0)} \sigma_j \sigma_k \rho_{jk}} \right), \;\;\; 1\leq i\leq N.
\end{equation}
Here, $Q$ is the orthogonal matrix of eigenvectors of $\Sigma$, and $\lambda$ the vector of eigenvalues.

The initial condition for all PDEs is given by $g(L)$ where $g$ is the payoff at time $T$.
The quantity of interest is $u^{(1,1)}(Z(0),T;\lambda)$, where $Z(0) = Q^T \ln(L(0)) + \beta(T)$.

\section{Implementation and numerical results}\label{sec:Num}
We study two types of derivatives to test the flexibility and accuracy of the approach and benchmark against
Monte Carlo results:
\begin{itemize}
\item
short- to long-running Bermudan swaptions, where the combination of high-dimensionality and early exercise presents challenges for PDE and MC methods;
\item
a ratchet floor, where the path-dependency is conceptually straightforward to include in a MC solver and needs adaptation of the PDE solver.
\end{itemize}

\subsection{Implementation of PDE solvers}

To compute the approximate solution defined by equation (\ref{eq:LMMPCA}) we need to numerically solve one- and two-dimensional PDEs of the type (\ref{onedpde}) and (\ref{twodpde}). 
These are standard and we used the following approach.


The computational domain is unbounded in the $z$-coordinates. To avoid the introduction of artificial boundary conditions necessary when localising the domain, 
 for each coordinate $z_i$,
we map the interval $(-\infty,\infty)$ to $(0,1)$ via
\begin{equation}\label{eq:CoordTrafoy}
	y_i = \frac{1}{\pi}\arctan\left(\gamma_i z_i + c_i\right)+ \frac{1}{2},
\end{equation}
with parameters $\gamma_i$ and $c_i$. Under a standard growth condition on the solution at infinity,
the resulting PDE is fully specified without boundary conditions at $z_i \in \{0,1\}$, because the resulting non-constant coefficients of the transformed diffusion-equation
vanish sufficiently fast at the boundaries (see \cite{RW07,ZL03}). For call-type options such as the Bermudan swaption discussed below we apply a payout cutoff at a value $g_{max} = 1000$, which does not significantly impact the computed option value.

We consider an equidistant grid with $J+1$ gridpoints along each axis, such that in original coordinates the mesh is denser in the interesting region, which depends on the LIBOR rates at $\tau=0$. For instance, in the case $L_i(0)=0.1$, which will be considered later, we choose $\gamma_i$ and $c_i$ such that LIBORs between $0.02$ and $0.5$ are mapped to the interval $[0.1,0.9]$.

For the discretisation we use the Crank-Nicolson scheme with central spatial differences.
In the two-dimensional case, we combine this with an Alternating Direction Implicit (ADI) factorisation \cite{MR55}, such that the resulting tridiagonal matrix systems can be solved efficiently in linear time
(i.e., proportional to the system size).
As the coefficients of the PDEs are constant in time, an initial LU factorisation of the tridiagonal matrices gave significant further speed-up.

Depending on the derivative contract, there can be additional parameters and interface conditions to be taken into account. The two examples we considered are Bermudan swaptions, which offer early exercise rights at discrete points in time, and Ratchet floors, where a strike parameter is reset depending on LIBORs at tenor dates, which makes the payoff strongly path-dependent. 
We describe both in more detail in the next sections.

All prices reported are relative to the bond at time $t = T_1 = 0$, i.e., in units of $P(T_1)$.

\subsection{Bermudan swaption}
\label{subsec:BS}

A Bermudan (payer) swaption with strike price $K$ can be exercised at any one of a set of exercise dates $\{T_{e_1},\ldots,T_{e_{N'}}\}\subseteq \{ T_1,\ldots,T_N\}$.
Here, we consider as an example 3-month LIBOR, i.e., $\alpha=0.25$ and $T_i=\alpha \, (i-1) = 0.25 \, (i-1)$ for $1\le i\le N$, and Bermudan swaptions which can be exercised yearly, i.e., $\{T_{e_1}\ldots,T_{e_{N'}}\} = \{T_1,T_5,T_9,\ldots,T_{N}\}$, assuming that $N-1$ can be divided by 4. If the Bermudan swaption is exercised at $T_i$ then the holder receives a (payer) swaption with payout
\begin{equation}
V_{Swaption,i} = \max\left(\alpha\cdot\sum_{j=i}^N [L_j(T_i)-K]/P(T_i,T_{j+1}),0\right) .
\end{equation}
The value $V_{BS,1}$ of a Bermudan swaption is thus determined by backward induction through $V_{BS,N} = V_{Swaption,N}$ and
\begin{equation}
\label{ex-cond}
V_{BS,i} = \max(V_{BS,i+4},V_{Swaption,i})  \text{ at } i\in \{1,5,\ldots,N-4\}.
\end{equation}
Between $T_i$ and $T_{i+4}$, $i\in\{1,5,\ldots,N-4\}$, the value function $V_{BS,i}$ satisfies the LMM PDE (\ref{eq:Feynman-Kac-PDE})--(\ref{eq:Feynman-Kac-PDEBoundary}),(\ref{eq:LIBORDynamics})--(\ref{eq:LIBORDrift_2}), which we approximate by PCA and first order anchored ANOVA decomposition as discussed in Section \ref{sec:PCALMM}.
The interface condition (\ref{ex-cond}) can easily be incorporated in the present PDE discretisation by evaluating (\ref{ex-cond}) on the computational grid.

As reference solutions for the PDE results we use Monte Carlo (MC) estimates. 
The numerical approximation of multi-dimensional American and Bermudan options by Monte Carlo methods is an area of active current research. We mention recent work
on the computation of tight bounds via iteration approaches (e.g., \cite{KS06}) and via pathwise optimisation (see \cite{DFM12}).
Here, we use the well-established and popular primal-dual approach for exercise policy learning due to Andersen and Broadie (see \cite{A04}),
which provides Monte Carlo estimates for a lower bound $V_{MC}^-$ and upper bound $V_{MC}^- + \Delta_0$ to the true option value.

In the simulations, we used $N_1 = 10^6$ paths for learning the exercise policy (of type `exercise strategy 1' in the notation of \cite{A04}), 
$N_2 = 10^7$ paths to calculate the lower bound and $N_{outer} = 5000$ and $N_{inner} = 1000$ paths for the outer and inner MC runs to compute the upper bound.
For the time discretization of the LMM SDEs we used the log-Euler scheme with $M_{MC}=5$ time steps per interval of length $\alpha$.
In the tests with ``frozen'' drift (i.e., lognormal LIBORs), the discretisation is exact for $M\ge 1$.
For the PDE we used $J=601$ grid points in every direction and the Crank-Nicolson scheme with $M_{PDE}=10$ time steps per time interval of length $\alpha$. 
The numerical parameters, summarised in Table \ref{tab:NumParams}, were chosen such that the numerical error is small compared to the difference between PDE and MC solution and is typically of order $0.1\%$ or less of the derivative value.

\begin{table}[tpb]
\begin{center}
\begin{tabular}{ccl}
Parameter & Value & Description \\ \hline\hline
$J$ & $601$ & Number of grid points in each direction  \\
$M_{PDE}$ & $10$ & Timesteps per interval of length $\alpha$ in the PDE computation \\
$N_1$ & $10^6$ & Number of MC paths for exercise policy learning \\
$N_2$ & $10^7$ & Number of MC paths to calculate the lower bound \\
$N_{outer}$ & $5000$ & Number of outer MC paths to calculate the upper bound \\
$N_{inner}$ & $1000$ & Number of inner MC paths to calculate the upper bound \\
$M_{MC}$ & $5$ & Timesteps per per interval of length $\alpha$ in the MC computation
\end{tabular}
\end{center}
\caption{Numerical parameters for the Bermudan swaption PDE and MC computations.}
\label{tab:NumParams}
\end{table}

\subsubsection*{Numerical results}

Results for Bermudan swaptions at-the-money (ATM, $K=0.1$) are shown in Table \ref{tab:ATM}. The PDE results are compared to the values calculated by MC simulation with frozen and full drift, to disentangle the effects of the drift approximation on the one hand and the dimension reduction on the other. The model parameters chosen were $\phi = 0.0413$ and $c=0.2$, with a flat initial LIBOR curve with $L_i(0)=0.1$, all identical to \cite{KS06}.

\begin{table}[tpb]
\begin{center}
\begin{tabular}{ccccccccccc}
$N$ & $V_{MC}^-$ & $\sigma$ & $V_{MC}^- + \Delta_0$ & $\sigma_{\Delta_0}$ & $V_{MC}^- + \Delta_0/2$ & $V_{PDE}$ & $\Delta_{abs}$ & $\Delta_{rel}$ \\ \hline\hline
5 & 1.75E-03 & 9.02E-07 & 1.75E-03 & 0.00E-00 & 1.75E-03 & 1.76E-03 & 1.18E-05 & 0.68\%  \\ 
11 & 1.21E-02 & 5.42E-06 & 1.22E-02 & 5.20E-06 & 1.21E-02 & 1.24E-02 & 2.61E-04 & 2.15\% \\ 
21 & 3.05E-02 & 1.14E-05 & 3.15E-02 & 4.65E-05 & 3.10E-02 & 3.14E-02 & 4.03E-04 & 1.30\% \\ 
41 & 6.17E-02 & 1.94E-05 & 6.68E-02 & 1.46E-04 & 6.42E-02 & 6.57E-02 & 1.44E-03 & 2.24\% \\ 
61 & 8.23E-02 & 2.29E-05 & 9.10E-02 & 2.16E-04 & 8.67E-02 & 9.04E-02 & 3.77E-03 & 4.35\% \\ 
81 & 9.45E-02 & 2.39E-05 & 1.06E-01 & 2.69E-04 & 1.00E-01 & 1.07E-01 & 6.84E-03 & 6.83\% \\ 
101 & 1.01E-01 & 2.38E-05 & 1.14E-01 & 3.82E-04 & 1.08E-01 & 1.18E-01 & 1.02E-02 & 9.45\% \\
\hline
5 & 1.75E-03 & 9.01E-07 & 1.75E-03 & 0.00E+00 & 1.75E-03 & 1.76E-03 & 1.33E-05 & 0.76\% \\ 
11 & 1.21E-02 & 5.43E-06 & 1.22E-02 & 5.16E-06 & 1.21E-02 & 1.24E-02 & 2.57E-04 & 2.11\% \\ 
21 & 3.06E-02 & 1.15E-05 & 3.16E-02 & 4.66E-05 & 3.11E-02 & 3.14E-02 & 3.17E-04 & 1.02\% \\ 
41 & 6.19E-02 & 1.98E-05 & 6.77E-02 & 1.63E-04 & 6.48E-02 & 6.57E-02 & 8.62E-04 & 1.33\% \\ 
61 & 8.26E-02 & 2.35E-05 & 9.27E-02 & 2.44E-04 & 8.77E-02 & 9.04E-02 & 2.74E-03 & 3.13\% \\ 
81 & 9.49E-02 & 2.45E-05 & 1.07E-01 & 2.85E-04 & 1.01E-01 & 1.07E-01 & 6.13E-03 & 6.07\% \\ 
101 & 1.02E-01 & 2.45E-05 & 1.16E-01 & 3.65E-04 & 1.09E-01 & 1.18E-01 & 9.23E-02 & 8.48\% \\
\end{tabular}
\end{center}
\caption{PDE results for ATM ($K=0.10$) Bermudan swaptions compared to frozen (top) and full (bottom) drift MC results. $V_{MC}^-$ and $\sigma$ are the lower MC bound and its estimated standard error. $V_{MC}^- + \Delta_0$ and $\sigma_{\Delta_0}$ are the upper MC bound and the estimated standard error of the MC spread $\Delta_0$. $V_{PDE}$ is the PDE result and the columns $\Delta_{abs}$ and $\Delta_{rel}$ show the absolute and relative difference to the best MC estimate $V_{MC}^- + \Delta_0/2$.}
\label{tab:ATM}
\end{table}

In these tests, the PDE method shows very good accuracy for up to $N=41$: $V_{PDE}$ is above $V_{MC}^-$ and below the upper MC bound in almost all cases for $N=21$ or $N=41$. For lower $N$ it is often slightly higher than the (in these cases fairly tight) upper bound, but the difference $\Delta_{abs}$  to the middle value $V_{MC}^-+\Delta_0/2$ -- which is considered to be a better estimate for the true price than both the lower or upper bound in \cite{KS06} -- is always less than $2.6$ basis points. 
For $N=61-101$, the PDE values are still close to the MC values. They are above the upper MC bound for $N=81$ and $N=101$, though. Taking into account the relatively large difference between lower and upper MC bounds, this seems to indicate that the applicability of the first-order, one-dimensional version of the PCA-ANOVA approach reaches its limits (in this setting) for problems with $N$ higher than $50-60$ (as does the MC approach used here).

In-the-money (ITM, $K=0.09$) and out-of-the-money (OTM, $K=0.11$) results are similar and are shown in Tables \ref{tab:ITM} and \ref{tab:OTM} in Appendix \ref{app:further}. As the overall value is largest for ITM and lowest for OTM options, the relative difference $\Delta_{rel} = \Delta_{abs}/(V_{MC}^-+\Delta_0/2)$ is typically smallest for ITM options and largest for OTM options.

\begin{figure}[tbp]
\begin{minipage}[l]{0.4\textwidth}
\begin{tikzpicture}[scale=.9]
	\begin{axis}[ylabel=$V_{BS}$ in bp, xlabel=$c$]
		\addplot+[only marks,mark=x,mark options={color=black,fill=black},mark size=3] coordinates { 
			(0.10,1.56E+02)
			(0.15,2.35E+02)
			(0.20,3.14E+02)
			(0.25,3.94E+02)
			(0.30,4.74E+02)
		};
		\addplot+[only marks,mark=-,mark options={color=black,fill=black},mark size=3] coordinates { 
			(0.10,1.53E+02)
			(0.15,2.30E+02)
			(0.20,3.06E+02)
			(0.25,3.81E+02)
			(0.30,4.56E+02)
		};
		\addplot+[only marks,mark=-,mark options={color=black,fill=black},mark size=3] coordinates { 
			(0.10,1.58E+02)
			(0.15,2.38E+02)
			(0.20,3.16E+02)
			(0.25,3.94E+02)
			(0.30,4.75E+02)
		};
	\end{axis}
\end{tikzpicture}
\end{minipage}
\begin{minipage}[c]{0.1\textwidth}
$\,$
\end{minipage}
\begin{minipage}[r]{0.4\textwidth}
\begin{tikzpicture}[scale=.9]
	\begin{axis}[ylabel=$V_{BS}$ in bp, xlabel=$c$]
		\addplot+[only marks,mark=x,mark options={color=black,fill=black},mark size=3] coordinates { 
			(0.10,3.24E+02)
			(0.15,4.89E+02)
			(0.20,6.57E+02)
			(0.25,8.29E+02)
			(0.30,10.1E+02)
		};
		\addplot+[only marks,mark=-,mark options={color=black,fill=black},mark size=3] coordinates { 
			(0.10,3.40E+02)
			(0.15,5.09E+02)
			(0.20,6.19E+02)
			(0.25,7.72E+02)
			(0.30,9.23E+02)
		};
		\addplot+[only marks,mark=-,mark options={color=black,fill=black},mark size=3] coordinates { 
			(0.10,3.10E+02)
			(0.15,4.65E+02)
			(0.20,6.77E+02)
			(0.25,8.43E+02)
			(0.30,10.1E+02)
		};
	\end{axis}
\end{tikzpicture}
\end{minipage}
\caption{PDE ($\scriptstyle \times$) and MC ($\scriptstyle -$) results for ATM Bermudan swaption values $V_{BS}$ for varying volatility $c$, given $\phi = 0.0413$ and $N=21$ (left) and $41$ (right). The difference between lower and upper MC bound is about $0.04 \, V_{BS}$ (left) and $0.10 \, V_{BS}$ (right). The standard deviation for the MC results is $\leq 0.0005 \, V_{BS}$ for the lower and $\leq 0.0025 \, V_{BS}$ for the upper bound. Note that all eigenvalues are proportional to $c^2$ and changing $c$ thus does not alter their relative sizes. 
}
\label{fig:varc}
\end{figure}
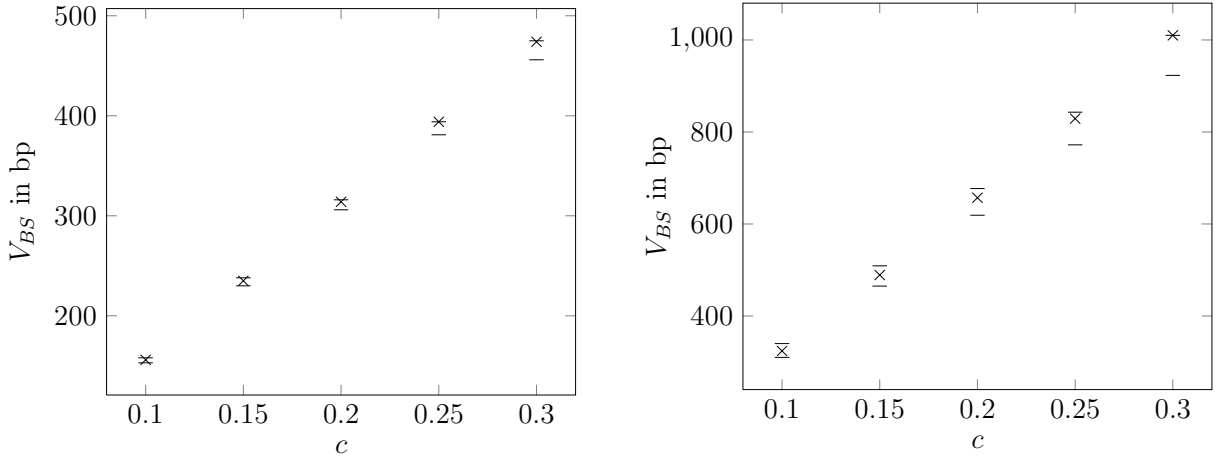

To assess the PCA-ANOVA approach under a range of market conditions, we present simulations for ATM Bermudan swaptions with differing parameters: Figures \ref{fig:varPhi} and \ref{fig:varc} show the results for varying correlation and volatility. For stronger correlation and lower volatility, where one would expect the highest accuracy, the PDE solution lies approximately in the middle between the lower and upper MC bound. For weaker correlation and higher volatility it tends towards and reaches the upper MC bound. Table \ref{tab:BSL002} shows that the PCA approach also performs well for a lower initial LIBOR curve with $L_i(0)=0.02$ for all $1\leq i\leq N$.

\begin{figure}[tbp]
\begin{minipage}[l]{0.4\textwidth}
\begin{tikzpicture}[scale=.9]
	\begin{axis}[ylabel=$V_{BS}$ in bp, scaled x ticks = false, xtick={1,2,3}, xlabel=$\phi/0.0413$, xticklabels={0.5,1,2}]
		\addplot+[only marks,mark=x,mark options={color=black,fill=black},mark size=3] coordinates { 
			(1,3.24E+02)
			(2,3.14E+02)
			(3,2.96E+02)
		};
		\addplot+[only marks,mark=-,mark options={color=black,fill=black},mark size=3] coordinates { 
			(1,3.18E+02)
			(2,3.06E+02)
			(3,2.85E+02)
		};
		\addplot+[only marks,mark=-,mark options={color=black,fill=black},mark size=3] coordinates { 
			(1,3.27E+02)
			(2,3.16E+02)
			(3,2.96E+02)
		};
	\end{axis}
\end{tikzpicture}
\end{minipage}
\begin{minipage}[c]{0.1\textwidth}
$\,$
\end{minipage}
\begin{minipage}[r]{0.4\textwidth}
\begin{tikzpicture}[scale=.9]
	\begin{axis}[ylabel=$V_{BS}$ in bp, scaled x ticks = false, xtick={1,2,3}, xlabel=$\phi/0.0413$, xticklabels={0.5,1,2}]
		\addplot+[only marks,mark=x,mark options={color=black,fill=black},mark size=3] coordinates { 
			(1,6.99E+02)
			(2,6.57E+02)
			(3,5.87E+02)
		};
		\addplot+[only marks,mark=-,mark options={color=black,fill=black},mark size=3] coordinates { 
			(1,7.28E+02)
			(2,6.19E+02)
			(3,5.96E+02)
		};
		\addplot+[only marks,mark=-,mark options={color=black,fill=black},mark size=3] coordinates { 
			(1,6.66E+02)
			(2,6.77E+02)
			(3,5.48E+02)
		};
	\end{axis}
\end{tikzpicture}
\end{minipage}
\caption{PDE ($\scriptstyle \times$) and MC ($\scriptstyle -$) results for ATM Bermudan swaption values $V_{BS}$ for varying correlation strength $\phi$ (where $\phi=0.0413$ is the value used in \cite{KS06}), given $c=0.2$ and $N=21$ (left) and $41$ (right). The difference between lower and upper MC bound is about $0.04 \, V_{BS}$ (left) and $0.10 \, V_{BS}$ (right). The standard deviation for the MC results is $\leq 0.0005 \, V_{BS}$ for the lower and $\leq 0.0025 \, V_{BS}$ for the upper bound. The corresponding eigenvalues are those shown in Figure \ref{fig:lambda11}.
}
\label{fig:varPhi}
\end{figure}
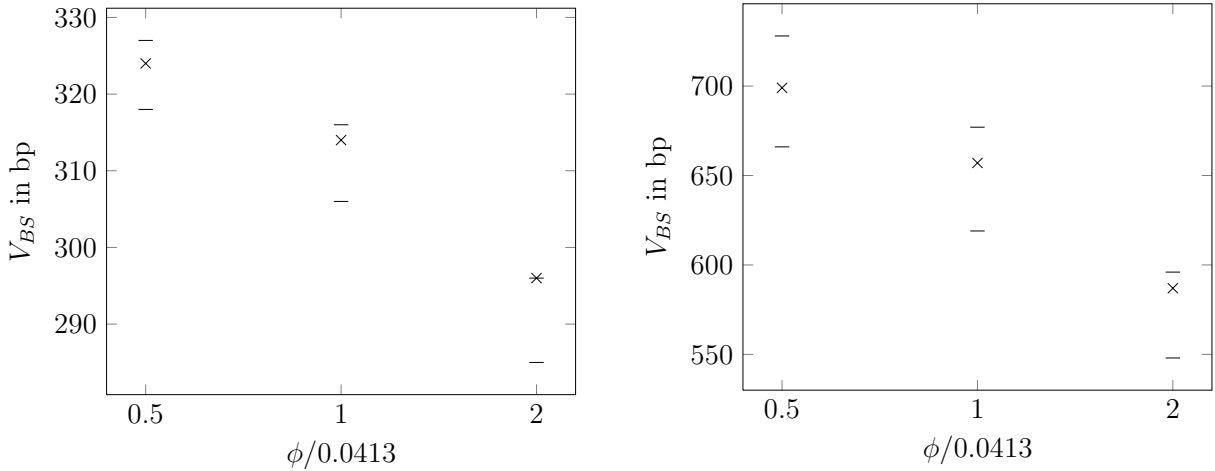

\begin{table}[tbp]
\begin{center}
\begin{tabular}{cccccccccc}
$N$ & $V_{MC}^-$ & $\sigma$ & $V_{MC}^- + \Delta_0$ & $\sigma_{\Delta_0}$ & $V_{MC}^- + \Delta_0/2$ & $V_{PDE}$ & $\Delta_{abs}$ & $\Delta_{rel}$ \\ \hline\hline
5 & 3.88E-04 & 2.02E-07 & 3.88E-04 & 0.00E+00 & 3.88E-04 & 3.89E-04 & 1.04E-06 & 0.27\% \\ 
11 & 2.86E-03 & 1.31E-06 & 2.87E-03 & 1.37E-06 & 2.87E-03 & 2.90E-03 & 3.78E-05 & 1.32\% \\ 
21 & 8.03E-03 & 3.19E-06 & 8.39E-03 & 1.51E-05 & 8.21E-03 & 8.17E-03 & -3.72E-05 & -0.45\% \\ 
41 & 2.00E-02 & 7.34E-06 & 2.32E-02 & 7.56E-05 & 2.16E-02 & 2.08E-02 & -7.83E-04 & -3.63\% \\ \hline
5 & 3.88E-04 & 2.02E-07 & 3.88E-04 & 0.00E+00 & 3.88E-04 & 3.89E-04 & 6.45E-07 & 0.17\% \\ 
11 & 2.86E-03 & 1.31E-06 & 2.88E-03 & 1.46E-06 & 2.87E-03 & 2.90E-03 & 3.74E-05 & 1.30\% \\ 
21 & 8.04E-03 & 3.18E-06 & 8.37E-03 & 1.46E-05 & 8.21E-03 & 8.17E-03 & -3.31E-05 & -0.40\% \\ 
41 & 2.01E-02 & 7.36E-06 & 2.30E-02 & 7.41E-05 & 2.15E-02 & 2.08E-02 & -7.03E-04 & -3.27\% \\ 
\end{tabular}
\end{center}
\caption{PDE results for ATM ($K=0.02$) Bermudan swaptions in a flat initial LIBOR curve setting with $L_i(0)=0.02 \; \forall 1\leq i\leq N$ compared to frozen (top) and full (bottom) drift MC results.}
\label{tab:BSL002}
\end{table}

\begin{figure}[tbp]
\centering
\begin{minipage}[l]{0.49\textwidth}
\begin{tikzpicture}[scale=0.9]
	\begin{axis}[xlabel=$k$, ylabel=$V_{PDE}$ in bp, ymode=normal, xmode=normal]
		\addplot+[only marks,mark=asterisk,mark options={color=blue,fill=blue},mark size=3] coordinates { 
		(	1	,	284.64	)
		(	2	,	303.67	)
		(	3	,	308.80	)
		(	4	,	310.82	)
		(	5	,	311.85	)
		(	6	,	312.37	)
		(	7	,	312.78	)
		(	8	,	313.06	)
		(	9	,	313.26	)
		(	10	,	313.41	)
		(	11	,	313.54	)
		(	12	,	313.64	)
		(	13	,	313.72	)
		(	14	,	313.81	)
		(	15	,	313.88	)
		(	16	,	313.96	)
		(	17	,	314.01	)
		(	18	,	314.07	)
		(	19	,	314.13	)
		(	20	,	314.18	)
		(	21	,	314.18	)
		
		};
		\addplot[black] coordinates
		{(1,316.167401) (21,316.167401)};
		\addplot[black] coordinates
		{(1,305.861356) (21,305.861356)};
	\end{axis}
\end{tikzpicture}	
\end{minipage}
\begin{minipage}[r]{0.49\textwidth}
\begin{tikzpicture}[scale=0.9]
	\begin{axis}[xlabel=$k$, ylabel=$|\Delta|$, ymode=log, xmode=log]
		\addplot+[only marks,mark=x,mark options={color=blue,fill=blue},mark size=3] coordinates { 
( 1 , 2.8464E-02)
(	2	,	1.9032E-03	)
(	3	,	5.1319E-04	)
(	4	,	2.0188E-04	)
(	5	,	1.0224E-04	)
(	6	,	5.2199E-05	)
(	7	,	4.1525E-05	)
(	8	,	2.7818E-05	)
(	9	,	1.9920E-05	)
(	10	,	1.5139E-05	)
(	11	,	1.2494E-05	)
(	12	,	9.8584E-06	)
(	13	,	8.9663E-06	)
(	14	,	8.2355E-06	)
(	15	,	7.6258E-06	)
(	16	,	7.2276E-06	)
(	17	,	5.7353E-06	)
(	18	,	5.8626E-06	)
(	19	,	5.5944E-06	)
(	20	,	5.2795E-06	)

		};
		\addplot+[only marks,mark=+,mark options={color=blue,fill=blue},mark size=3] coordinates { 
( 1 , 3.1418E-02)
(	2	,	2.9540E-03	)
(	3	,	1.0508E-03	)
(	4	,	5.3760E-04	)
(	5	,	3.3572E-04	)
(	6	,	2.3348E-04	)
(	7	,	1.8128E-04	)
(	8	,	1.3976E-04	)
(	9	,	1.1194E-04	)
(	10	,	9.2018E-05	)
(	11	,	7.6879E-05	)
(	12	,	6.4385E-05	)
(	13	,	5.4527E-05	)
(	14	,	4.5561E-05	)
(	15	,	3.7325E-05	)
(	16	,	2.9699E-05	)
(	17	,	2.2472E-05	)
(	18	,	1.6737E-05	)
(	19	,	1.0874E-05	)
(	20	,	5.2795E-06	)

		};
	\end{axis}
\end{tikzpicture}
\end{minipage}
\begin{tabular}{ccccccc}
$k$ & 1 & 2 & 3 & 6 & 11 & 21 \\ \hline\hline
$V_{PDE}$ & 285 bp & 304 bp& 309 bp& 312 bp & 314 bp& 314 bp\\
$\Delta$ & -9.40\% & -3.34\% & -1.71\% & -0.58\% & -0.20\% & 0.00\% \\
\end{tabular}
$\,$\vspace{3em}
\begin{minipage}[l]{0.49\textwidth}
\begin{tikzpicture}[scale=0.9]
	\begin{axis}[xlabel=$k$, ylabel=$V_{PDE}$ in bp, ymode=normal, xmode=normal]
		\addplot+[only marks,mark=asterisk,mark options={color=blue,fill=blue},mark size=3] coordinates { 
		(	1	,	536.76	)
		(	2	,	607.76	)
		(	3	,	629.22	)
		(	4	,	638.62	)
		(	5	,	643.56	)
		(	6	,	646.51	)
		(	7	,	648.45	)
		(	8	,	649.82	)
		(	9	,	650.83	)
		(	10	,	651.62	)
		(	11	,	652.18	)
		(	12	,	652.69	)
		(	13	,	653.11	)
		(	14	,	653.46	)
		(	15	,	653.76	)
		(	16	,	654.02	)
		(	17	,	654.25	)
		(	18	,	654.45	)
		(	19	,	654.63	)
		(	20	,	654.79	)
		(	21	,	654.94	)
		(	22	,	655.07	)
		(	23	,	655.20	)
		(	24	,	655.32	)
		(	25	,	655.43	)
		(	26	,	655.54	)
		(	27	,	655.64	)
		(	28	,	655.74	)
		(	29	,	655.83	)
		(	30	,	655.92	)
		(	31	,	656.01	)
		(	32	,	656.09	)
		(	33	,	656.17	)
		(	34	,	656.25	)
		(	35	,	656.32	)
		(	36	,	656.40	)
		(	37	,	656.47	)
		(	38	,	656.54	)
		(	39	,	656.61	)
		(	40	,	656.68	)
		(	41	,	656.68	)
		
		};
		\addplot[black] coordinates
		{(1,677.460912) (41,677.460912)};
		\addplot[black] coordinates
		{(1,618.671085) (41,618.671085)};
	\end{axis}
\end{tikzpicture}
\end{minipage}
\begin{minipage}[l]{0.49\textwidth}
\begin{tikzpicture}[scale=0.9]
	\begin{axis}[xlabel=$k$, ylabel=$|\Delta|$, ymode=log, xmode=log]
		\addplot+[only marks,mark=x,mark options={color=blue,fill=blue},mark size=3] coordinates { 
( 1 , 5.3676E-02)		
(	2	,	7.0999E-03	)
(	3	,	2.1465E-03	)
(	4	,	9.3977E-04	)
(	5	,	4.9344E-04	)
(	6	,	2.9542E-04	)
(	7	,	1.9388E-04	)
(	8	,	1.3668E-04	)
(	9	,	1.0158E-04	)
(	10	,	7.8393E-05	)
(	11	,	5.6175E-05	)
(	12	,	5.0943E-05	)
(	13	,	4.2386E-05	)
(	14	,	3.5307E-05	)
(	15	,	2.9937E-05	)
(	16	,	2.5690E-05	)
(	17	,	2.2567E-05	)
(	18	,	2.0022E-05	)
(	19	,	1.7939E-05	)
(	20	,	1.6165E-05	)
(	21	,	1.4966E-05	)
(	22	,	1.3295E-05	)
(	23	,	1.2595E-05	)
(	24	,	1.1973E-05	)
(	25	,	1.1352E-05	)
(	26	,	1.0746E-05	)
(	27	,	1.0101E-05	)
(	28	,	9.7195E-06	)
(	29	,	9.3857E-06	)
(	30	,	9.0992E-06	)
(	31	,	8.8772E-06	)
(	32	,	7.9134E-06	)
(	33	,	8.0213E-06	)
(	34	,	7.8313E-06	)
(	35	,	7.6643E-06	)
(	36	,	7.3851E-06	)
(	37	,	7.2644E-06	)
(	38	,	7.2035E-06	)
(	39	,	7.0324E-06	)
(	40	,	7.0141E-06	)

		};
	\addplot+[only marks,mark=+,mark options={color=blue,fill=blue},mark size=3] coordinates { 
( 1 , 6.5668E-02)
(	2	,	1.1992E-02	)
(	3	,	4.8922E-03	)
(	4	,	2.7457E-03	)
(	5	,	1.8060E-03	)
(	6	,	1.3125E-03	)
(	7	,	1.0171E-03	)
(	8	,	8.2322E-04	)
(	9	,	6.8655E-04	)
(	10	,	5.8496E-04	)
(	11	,	5.0657E-04	)
(	12	,	4.5040E-04	)
(	13	,	3.9945E-04	)
(	14	,	3.5707E-04	)
(	15	,	3.2176E-04	)
(	16	,	2.9182E-04	)
(	17	,	2.6613E-04	)
(	18	,	2.4357E-04	)
(	19	,	2.2354E-04	)
(	20	,	2.0561E-04	)
(	21	,	1.8944E-04	)
(	22	,	1.7447E-04	)
(	23	,	1.6118E-04	)
(	24	,	1.4858E-04	)
(	25	,	1.3661E-04	)
(	26	,	1.2526E-04	)
(	27	,	1.1451E-04	)
(	28	,	1.0441E-04	)
(	29	,	9.4692E-05	)
(	30	,	8.5306E-05	)
(	31	,	7.6207E-05	)
(	32	,	6.7330E-05	)
(	33	,	5.9416E-05	)
(	34	,	5.1395E-05	)
(	35	,	4.3564E-05	)
(	36	,	3.5899E-05	)
(	37	,	2.8514E-05	)
(	38	,	2.1250E-05	)
(	39	,	1.4046E-05	)
(	40	,	7.0141E-06	)
	};
	\end{axis}
\end{tikzpicture}
\end{minipage}
\begin{tabular}{cccccccc}
		$k$ & 1 & 2 & 3 & 6 & 11 & 21 & 41 \\ \hline\hline
		$V_{PDE}$ & 537 bp & 608 bp & 629 bp & 647 bp & 652 bp & 655 bp & 657 bp \\
		$\Delta$ & -18.26\% & -7.45\% & -4.18\% & -1.55\% & -0.69\% & -0.27\% & 0.00\% \\
\end{tabular}
$\,$\vspace{3em}
\caption{
Approximations $V_{PDE}$ to the value of the ATM Bermudan swaption with $N=21$ (top) and $N=41$ (bottom) when including only the terms up to $i=k$ in equation (\ref{eq:LMMPCA}) (left). The solid horizontal lines are the lower and upper MC bounds. The graphs on the right show the terms in (\ref{eq:LMMPCA}) for each individual $i$ ($\scriptstyle \times$) and the sum of the remaining terms $i=k+1,\ldots,N$ ($\scriptstyle +$).
The tables give numerical values, with the row $\Delta$ showing the relative difference to the PDE result with all $N$ terms.}
\label{fig:dimconv}
\end{figure}

Finally, Figure \ref{fig:dimconv} shows how the value of the PDE approximation changes when we consider in
(\ref{eq:LMMPCA}) only
the 1-dimensional PDE solution and the 2-dimensional PDE solutions associated with the $k$ largest eigenvalues. Specifically, the case $k=1$ is the one-dimensional approximation using the first principal component, and the case
$k=2$ the standard two-dimensional PCA approximation. The contributions for different eigenvalues approximately lie on a line with slope $-2$ in the log-log-plots in Figure \ref{fig:dimconv}, just as the eigenvalues in Figure \ref{fig:lambda11}, i.e., they show the same decay. Since the $i$-th contribution is equal to $\lambda_i \frac{\partial u}{\partial \lambda_{i}}$, this suggests that the partial derivatives $\frac{\partial u}{\partial \lambda_{i}}$ are all of similar size.

Evidently, it is in fact necessary to include the contributions from several of the largest eigenvalues to compute an accurate solution. At the same time, the solution levels out after including about $10$ dimension. This is in line with the decay of the eigenvalues and the fact that the payoff in this case is almost parallel to the eigenvector of the first dimension. For models where the eigenvalues of the covariance matrix decay fast enough -- which includes many models in mathematical finance -- the PCA-ANOVA PDE approach might be used with a fixed number $k$ for a wide range of values $N$. This further reduces the computational effort -- which in the given case is roughly proportional to $k$ -- without sacrificing significant accuracy.

\subsubsection*{Run times}

While considerable effort went into the efficient implementation of both the PDE and MC methods, 
there is still room for performance improvement, for instance on the algorithmic level and in the numerical parameter choices.
Thus we only want to comment on approximate run times: For the PDE calculations, the computation times for $N=5,11,21$ and $41$ were on the order of $10$, $60$, $240$ and $1000$ seconds, resp., using Matlab on a AMD Phenom(tm) II X4 925 Processor (2.8 GHz) with 3.8 GB RAM. 
The run time is roughly quadratic in $N$, because the number of PDE solutions required to evaluate (\ref{eq:LMMPCA}) is $N$, and the expiry is $T=N/4$, so the number of (Crank-Nicolson) time steps with fixed step size, for a given PDE, is proportional to $N$.

For the MC simulations the corresponding computation times were of order $250+65$, $1100+600$, $4000+5200$ and $17000+44000$ seconds, resp., for frozen drift and $250+65$, $1200+700$, $4500+5700$ and $20000+54000$, resp., for full drift. Here the first number is the computation time for the lower bound and the second number is the additional time necessary to compute the upper bound.
The run time is roughly quadratic in $N$ for the lower bound, since $N$ processes have to be simulated over $N$ tenor dates, and roughly cubic in $N$ for the upper bound.

Despite the approximate nature of these computation times, it becomes clear that the PDE method is not only competitive time-wise, but indeed faster by a factor of $25-70$ in our implementation. 
To get an optimal allocation of computational resources,
 one could also try to further optimise the relative size of the numerical parameters, such that the computation time is optimal for a given size of the combined discretisation error.
By, say, halving the mesh size in the two directions of the computational grid one quadruples the computational time,
while there is no practical accuracy gain in bringing the discretisation error substantially below the error of the dimension reduction. A similar statement is true for the Monte Carlo estimators. A precise comparison of efficiency is therefore delicate.

Both the PDE and MC methods used in this section are limited in their accuracy: the MC method uses a class of exercise strategies which generally does not include the optimal one; the PDE method employs a drift approximation and asymptotic expansion. For a wide range of $N$, the errors are comparable. Possible accuracy improvements on the basis of the decomposition are outlined in Sections \ref{sec:anova} and \ref{sec:Con}, and will be tested on the following application.

\subsection{Ratchet floor}
\label{subsec:ratfloor}

A Ratchet floor with strike price $K_1$ and parameters $a,b,c$ is a portfolio (sum) of floorlets with payouts
$\max\left\{K_i-L_i(T_{i}),0\right\}$ at $T_{i+1}$, where the strike prices $K_i$ are recursively determined from the given initial strike $K_1$ by 
\begin{eqnarray}
K_{i} &=& \max\left(aL_{i-1}(T_{i-1})+bK_{i-1}+c,0\right), \quad i>1,
\label{reset}
\end{eqnarray}
see \cite{PSV11}.
The (relative) price of the Ratchet floor for $t=0$ is given by
\begin{equation}
	V_{RF}(0) = \sum_{i=1}^N E\left[V_{Fl,i}(T_i)/\mathcal{N}(T_i) | \mathcal{F}_0\right],
\end{equation}
where $V_{Fl,i}$ is the value of the $i$-th floorlet.
Due to the linearity in the sum on the right-hand side of this equation it is sufficient to be able to calculate the price of a single floorlet. Without loss of generality we will thus focus on $V_{Fl,N}$.\\
The ratchet feature (\ref{reset}) makes the problem high-dimensional and strongly path-dependent as the payoff depends on the values of all $L_i$ at different points in time.
To solve the problem by a backward equation, we need to make it Markovian by including the strike dynamics with the evolution of the LIBORs, and to specify the value function as a function of all the above.
As the strike changes discretely in time, the value function satisfies the standard LMM PDE between the $T_i$.
At each tenor time $T_i$, the jump condition
\begin{equation}
\label{eq:jumpcond}
V_{Fl,i}(T_i-,K,L_1,\ldots,L_N) = V_{Fl,i}(T_i+,\max\left(a L_{i-1}+b K+c,0\right)\!, L_1,\ldots,L_N)
\end{equation}
holds.
Here, $T_i+$ denotes the limit coming from larger $t$ where the solution is already computed, and we use this to compute the solution just prior to $T_i$ before the strike is updated.
Details of the complete PDE model and its mathematical analysis can be found in \cite{PSV11}.

We approximate the solution on a grid in $K$-direction, and
compute the updated solution for each grid point via cubic spline interpolation for the corresponding value of $K_{i+1}$.
This adds an extra dimension $N+1$ to the problem, and effectively the one-dimensional ANOVA terms (corresponding to $z_1$) now live on a two-dimensional grid, and the two-dimensional ANOVA terms (corresponding to $z_1$
and an additional $z_i$) on a three-dimensional grid, formulaically (cf.\ Section \ref{sec:anova}),
\begin{eqnarray}
u^{(2,1)}(K,z) &=& u_0^{(2)}(K_1,Z(0);K,z_1) + \sum_{i=2}^N u_i^{(2)}(K_1,Z(0);K,z_1;z_i),
\end{eqnarray}
where the superscript `2' stands for the variables $K$ and $z_1$ and the `1' for the extra coordinate $z_i$ in the expansion.
It is conceivable to use an expansion in direction $K$ with anchor $K_1$ of the form (\ref{anovars}), which becomes
\begin{eqnarray}
u^{(1,1)}(K,z) = u_0^{(1)}(K_1,Z(0);z_1)  + \sum_{i=2}^N u_i^{(1)}(K_1,Z(0);z_1;z_i) + u_{N+1}^{(1)}(K_1,Z(0);z_1;K),
\end{eqnarray}
where the first superscript `1' stands for $z_1$ and the second `1' for the expansion in one coordinate $z_i$ or $K$.
In other words, we identify the path-dependent quantity $K$ defined in (\ref{reset}) with an extra dimension $z_{N+1}$ in (\ref{anovars}) with $N$ replaced by $N+1$.
We do not pursue this further here.

Due to the smoothness of the solution in the $K$-direction and the higher order of the spline interpolation relative to the finite difference scheme in $L$ directions, a relatively coarse mesh on the $K$ axis is sufficient. Specifically, we use 21, 41 and 21 spline nodes for the intervals $[0,0.05]$, $[0.05,0.15]$ and $[0.15,K_{max}]$, resp., with $K_{max} = 0.5$, to give a total of 81 nodes on the $K$-axis.
In comparison, we use $J = 401$ grid points in the $z_i$-directions and the Crank-Nicolson scheme with $M=10$ time steps per interval of length $\alpha=0.25$ between tenors. The numerical parameters are again chosen such that the numerical error is small compared to the difference between PDE and MC solution and is typically around $0.1\%$ or less of the derivative value.
The model parameters chosen were $\phi = 0.0413$ and $c=0.2$, with a flat initial LIBOR curve with $L_i(0)=0.1$, identical to those for the Bermudan swaptions in Table \ref{tab:ATM}.

We again use MC estimates as reference solutions. For a path-dependent option without early-exercise features like the Ratchet floor we can use a straightforward MC calculation.
In the tests, we sampled $N_1 = 10^7$ paths; for the time discretization, we used the log-Euler scheme with $M=10$ time steps per interval of length $\alpha$.

\begin{table}[tpb]
\begin{center}
\begin{tabular}{ccl}
Parameter & Value & Description \\ \hline\hline
$J$ & $401$ & Number of grid points in each LIBOR direction  \\
$I$ & $81$ & Number of grid points in strike price direction \\
$M_{PDE}$ & $10$ & Timesteps per interval of length $\alpha$ in the PDE computation \\
$N_1$ & $10^7$ & Number of MC paths \\
$M_{MC}$ & $10$ & Timesteps per per interval of length $\alpha$ in the MC computation
\end{tabular}
\end{center}
\caption{Numerical parameters for the Ratchet floor PDE and MC computations.}
\label{tab:NumParamsRF}
\end{table}

\subsubsection*{Numerical results}

The numerical results for the Ratchet floors are shown in Table \ref{tab:RFf} for $N=5,11,21$. We consider three different configurations $(a,b,c)$ and three different strike prices $K = 0.1, 0.11, 0.09$ (ATM, ITM and OTM, resp.). The absolute difference between the PDE and MC solution is never more than $1.14$ basis points and the relative difference is below $1\%$ in all but one case.

\begin{table}[htbp]
\begin{center}
\begin{tabular}{ccccccc}
$(a,b,c)$ & $N$ & $V_{MC}$ & $\sigma$ & $V_{PDE}$ & $\Delta_{abs}$ & $\Delta_{rel}$ \\ \hline\hline
	$K_1 = 0.10$ &&&&&&\\
0/1/0 & 5 & 7.08E-003 & 2.95E-006 & 7.04E-003 & -3.39E-05 & -0.48\% \\ 
 & 11 & 9.65E-003 & 3.87E-006 & 9.61E-003 & -3.87E-05 & -0.40\% \\ 
 & 21 & 1.07E-002 & 4.20E-006 & 1.07E-002 & 2.98E-05 & 0.28\% \\ 
0.2/0.9/0 & 5 & 3.06E-002 & 4.23E-006 & 3.06E-002 & -6.30E-05 & -0.21\% \\ 
 & 11 & 4.94E-002 & 4.74E-006 & 4.93E-002 & -9.12E-05 & -0.18\% \\ 
 & 21 & 5.10E-002 & 5.02E-006 & 5.09E-002 & -3.33E-05 & -0.07\% \\ 
0.25/0.95/-0.01 & 5 & 3.29E-002 & 4.03E-006 & 3.29E-002 & -5.74E-05 & -0.17\% \\ 
 & 11 & 6.06E-002 & 5.13E-006 & 6.06E-002 & -1.26E-05 & -0.02\% \\ 
 & 21 & 7.36E-002 & 8.13E-006 & 7.36E-002 & -1.13E-05 & -0.02\% \\ 
\hline
$K_1 = 0.11$ &&&&&&\\
	0/1/0 & 5 & 1.27E-002 & 3.88E-006 & 1.26E-002 & -5.34E-05 & -0.42\% \\ 
	 & 11 & 1.44E-002 & 4.74E-006 & 1.43E-002 & -7.10E-05 & -0.49\% \\ 
	 & 21 & 1.44E-002 & 4.94E-006 & 1.44E-002 & -5.28E-05 & -0.37\% \\ 
	0.2/0.9/0 & 5 & 3.63E-002 & 4.33E-006 & 3.63E-002 & -6.40E-05 & -0.18\% \\ 
	 & 11 & 5.20E-002 & 4.76E-006 & 5.20E-002 & -9.04E-05 & -0.17\% \\ 
	 & 21 & 5.17E-002 & 5.02E-006 & 5.17E-002 & -3.94E-05 & -0.08\% \\ 
	0.25/0.95/-0.01 & 5 & 4.00E-002 & 4.11E-006 & 4.00E-002 & -5.81E-05 & -0.15\% \\ 
	 & 11 & 6.52E-002 & 5.12E-006 & 6.52E-002 & -1.75E-05 & -0.03\% \\ 
	 & 21 & 7.58E-002 & 8.08E-006 & 7.57E-002 & -3.69E-05 & -0.05\% \\ 
\hline	
$K_1 = 0.09$ &&&&&&\\
	0/1/0 & 5 & 3.20E-003 & 1.94E-006 & 3.17E-003 & -2.32E-05 & -0.73\% \\ 
	 & 11 & 5.83E-003 & 2.95E-006 & 5.82E-003 & -1.56E-05 & -0.27\% \\ 
	 & 21 & 7.39E-003 & 3.42E-006 & 7.47E-003 & 7.68E-05 & 1.04\% \\ 
	0.2/0.9/0 & 5 & 2.51E-002 & 4.07E-006 & 2.50E-002 & -7.27E-05 & -0.29\% \\ 
	 & 11 & 4.68E-002 & 4.71E-006 & 4.67E-002 & -1.02E-04 & -0.22\% \\ 
	 & 21 & 5.03E-002 & 5.02E-006 & 5.02E-002 & -4.86E-05 & -0.10\% \\ 
	0.25/0.95/-0.01 & 5 & 2.59E-002 & 3.88E-006 & 2.58E-002 & -6.73E-05 & -0.26\% \\ 
	 & 11 & 5.61E-002 & 5.14E-006 & 5.60E-002 & -2.07E-05 & -0.04\% \\ 
	 & 21 & 7.15E-002 & 8.19E-006 & 7.15E-002 & -1.61E-05 & -0.02\% \\ 
\end{tabular}
\end{center}
\caption{PDE results for Ratchet floors compared to full drift MC results. The column $V_{PDE}$ shows the computed PDE value. Columns $\Delta_{abs}$ and $\Delta_{rel}$ show the absolute and relative difference to the MC estimate $V_{MC}$. PDE results compared to frozen drift MC results are shown in Table \ref{tab:RF} in the Appendix.}
\label{tab:RFf}
\end{table}

\subsubsection*{Run times}

We again report approximate computation times with the same caution as in the previous section. For $N=5,11$ and $21$ the PDE run times were of order $400$, $2400$ and $10000$ seconds, resp.,  on a AMD Phenom(tm) II X4 925 Processor (2.8 GHz) with 3.8 GB RAM. For frozen drift the corresponding MC run times were of order $160$, $900$ and $3500$ seconds, while for full drift they werde $190$, $1100$ and $4600$ seconds. The MC computation was faster by a factor of $2$ to $3$ compared to the PDE computation. Both were roughly quadratic in $N$.
The MC simulation also permits the computation of values for multiple parameters $(a,b,c)$ in parallel, with only a small increase in computation time.
Given the fast decay of the correction terms in the ANOVA decomposition, as illustrated in Figure \ref{fig:dimconv} for the Bermudan swaption, it would be possible to compute only the first $k\ll N$ ANOVA terms without significant loss of accuracy, which brings the computational times for the PDE in the range of, or below, the MC ones.

\subsection{Higher-order Taylor terms}
\label{subsec:higher-order}

In Sections \ref{subsec:taylor} and \ref{sec:anova} we explained how to include higher order terms in the Taylor and ANOVA expansions. This extension is expected to decrease the size of the error to correspondingly higher orders of lambda. In this section, we present experimental results demonstrating empirically the validity of this assertion. We also investigate the behaviour of the approximation when including fully the first $r$ eigenvalues, i.e., expanding around
\begin{eqnarray}
\label{lam0}
\lambda^0 = (\lambda_1,\ldots,\lambda_r,0,\ldots).
\end{eqnarray}

Previous investigations in \cite{RW07} and \cite{SGW12} have considered examples with equity baskets, for $(r,s) = (1,1)$, $(1,2)$ and $(2,1)$, where 
$s$ is the order of the Taylor expansions.
See also the discussion at the end of Section \ref{sec:anova}.
Here, we systematically explore the impact of varying $r$ and $s$. 
\begin{table}[tbp]
		\begin{center}
			\begin{tabular}{crc|crc}
				& $V$ & $\sigma$ & & $V$ & $\sigma$ \\ \hline
				$V_{100}$ & 498.902 & 0.002 & $V_{222}$ & 0.111 & 0.002 \\
				$V_{111}$ & -5.079 & 0.002 & $V_{300}$ & 494.610 & 0.010 \\ 
				$V_{112}$ & -5.208 & 0.003 & $V_{311}$ & -0.528 & 0.004 \\ 
				$V_{113}$ & -5.179 & 0.003 & $V_{322}$ &  0.033 & 0.001 \\ 
				$V_{122}$ & 0.250 & 0.002 & $V_{400}$ & 494.233 & 0.010 \\
				$V_{123}$ & 0.214 & 0.003 & $V_{411}$ & -0.129 & 0.003 \\ 
				$V_{133}$ & 0.109 & 0.010 & $V_{500}$ & 494.071 & 0.010 \\ 
				$V_{200}$ & 495.657 & 0.009 & $V_{511}$ & 0.041 & 0.002 \\
				$V_{211}$ & -1.644 & 0.006 & $V_{full}$ & 494.107 & 0.003 \\
			\end{tabular}
		\end{center}
	\caption{Numerical results (in bp) for the different terms $V_{rst}$ as in (\ref{Vrst}), for a Ratchet floor with $(a,b,c)=(0.2,0.9,0)$ and $N=11$. $r$ is the number of fully included eigenvalues, $s$ is the order of the Taylor expansion and $t$ is the order of the finite difference stencils used to compute the derivatives. The computations used $2\cdot 10^9$ MC paths for $V_{100}$, $V_{111}$, $V_{112}$, $V_{113}$ and $V_{full}$ and $2\cdot 10^8$ MC paths for all other values, where $\sigma$ is the root-mean-squared error.}
	\label{tab:RFhigherorder2}
\end{table}

Table \ref{tab:RFhigherorder2} shows estimates for the higher order terms in (\ref{eq:Extension1})
for a Ratchet floor with $(a,b,c)=(0.2,0.9,0)$ and $N=11$.
Precisely, for an $N-r$ dimensional multi-index $\omega = (\omega_1,\ldots,\omega_{N-r})$, 
we study, for different values of $r$, $s$ and $t$,
\begin{eqnarray}
\label{Vrst}
V_{rst} = V_{rst}(z_0,0;\lambda,\lambda^0) &=& \sum_{|\omega|=s} \Delta^{r,\omega,t}(u;z_0,0;\lambda,\lambda^0) \\
&\hspace{-4cm} =&\hspace{-2cm}
\sum_{|\omega|=s}
\frac{\lambda_{r+1}^{\omega_1} \cdot \ldots \cdot \lambda_{N}^{\omega_{N-r}}}{\omega!} \frac{\partial^{|\omega|} u}{\partial \lambda_{r+1}^{\omega_1}\ldots\partial \lambda_{N}^{\omega_{r}}}(z_0,0;\lambda^0) +
 O(\|\lambda^0-\lambda\|^{t+1}),
 \label{Vrst-order}
\end{eqnarray}
where $\Delta^{r,\omega,t}(u;z_0,0;\lambda,\lambda^0)$ 
is an approximation of order $t+1$ to the relevant mixed partial derivative term of $u$ at $(z_0,0)$ and $\lambda^0$ (see Section \ref{subsec:taylor}).
We have chosen finite difference stencils with weights as shown in Table \ref{tab:stencils}; for instance, the stencil in the second line is the standard right-sided difference and the one in the fifth line a standard second difference shifted to the right. 
Each of these finite differences requires the solution of $r+s$-dimensional problems, as detailed in
Section \ref{subsec:taylor}.
This computation was performed with a Monte Carlo method for illustration purposes and we comment on this at the end of this section.

\begin{table}[tbp]
	\centering
	\begin{tabular}{cccc|rc}
		\multicolumn{1}{c}{$u(3\lambda)$} & \multicolumn{1}{c}{$u(2\lambda)$} & \multicolumn{1}{c}{$u(\lambda)$} & \multicolumn{1}{c}{$u(0)$} & Taylor  term & $t+1$ \\ \hline
		 & & & $^1/_1$ & $u(0)$ & $-$ \\
		 & & $^1/_1$ & $-^1/_1$ & $\lambda u'(0)$ & 2\\
		 & $-^1/_2$ & $^4/_2$ & $- ^3/_2$ & $\lambda u'(0)$ & 3\\
		 $ ^2/_6$ & $- ^9/_6$ & $^{18}/_6$ & $- ^{11}/_6$ & $\lambda u'(0)$ & 4\\
		 & $ ^1/_2$ & $-^2/_2$ & $ ^1/_2$ & $\lambda^2u''(0)/2$ & 3 \\
		 $- ^1/_2$ & $^4/_2$ & $- ^5/_2$ & $^2/_2$ & $\lambda^2 u''(0)/2$ & 4 \\
		 $ ^1/_6$ & $- ^3/_6$ & $ ^3/_6$ & $- ^1/_6$ & $\lambda^3 u'''(0)/6$ & 4\\
	\end{tabular}
	\caption{Coefficients of the finite difference stencils used to approximate the Taylor expansion terms of a function $u(\lambda)$ at 0, where $t+1$ is the order of the approximation, see (\ref{Vrst-order}). Multi-dimensional stencils for partial derivatives were constructed by straightforward multiplication of the single-dimensional stencils.}
	\label{tab:stencils}
\end{table}

The magnitude of the $V_{rst}$ decreases significantly with increasing order $s$, as one would expect from the Taylor expansion. Likewise, the correction terms for higher $r$ are much smaller than for lower $r$. For example, $V_{311}$ is only about one tenth the size of $V_{111}$. This is again in line with the theoretical prediction, because $V_{300}$ should be closer to the full solution than $V_{100}$.

\begin{table}[tbp]
		\begin{center}
			\begin{tabular}{c|rrrr|rrrr}
				$r$ \textbackslash{} $s$ & \multicolumn{1}{c}{0} & \multicolumn{1}{c}{1} & \multicolumn{1}{c}{2} & \multicolumn{1}{c}{3} & \multicolumn{1}{c}{0} & \multicolumn{1}{c}{1} 
				& \multicolumn{1}{c}{2} & \multicolumn{1}{c}{3} \\ \hline
				1 & 4.795 & -0.284 & -0.034 & 0.075 & 0.970\% & -0.058\% & -0.007\% & 0.015\% \\
				2 & 1.550 & -0.093 & 0.018 & & 0.314\% & -0.019\%  & 0.004\% & \\
				3 & 0.502 & -0.026 & 0.007 & & 0.102\% & -0.005\% & 0.001\% & \\
				4 & 0.126 & -0.003 & & & 0.025\% & -0.001\% & & \\
				5 & -0.036 & 0.005 & & & -0.007\% & 0.001\% & & \\
			\end{tabular}
		\end{center}
	\caption{Absolute (left, in bp) and relative (right) difference between the PDE expansion approximation values $\sum_{i=0}^s V_{rii}$ and the full solution $V_{full}$ for different numbers $r$ of fully included eigenvalues and orders $s$ of the Taylor expansion. The numbers are based on the results in Table \ref{tab:RFhigherorder2}. The standard deviation of these estimates is $0.01$\,bp and $0.002\%$, resp.}
	\label{tab:RFhigherorder}
\end{table}

To analyse how the accuracy of the overall approximation changes with increasing $r$ and $s$, Table \ref{tab:RFhigherorder} shows the difference between the sum of all terms up to order $s$ and the full solution, i.e.,
\begin{equation}
\Delta_{abs}^{r,s} = \sum_{i=0}^s V_{rii} - V_{full}
\end{equation}
and
\begin{equation}
\Delta_{rel}^{r,s} = \frac{\sum_{i=0}^s V_{rii} - V_{full}}{V_{full}},
\end{equation}
for a range of values of $r$ and $s$. Again, the error decreases rapidly with increasing $r$ and $s$. The absolute error is well below 1\,bp for all 1st-order cases and well below 0.1\,bp for all 2nd-order cases. In general, the results suggest that it is possible to make the error negligibly small even with low to moderate values of $r$ and $s$. The computational effort is dominated by the term with highest $r+s$, which requires the computation of $O((N-r)^s)$ $r+s$-dimensional partial derivatives with stencils of order $s$.

A notable outlier is the case $(r,s) = (1,3)$, which is somewhat less accurate than for $(r,s)=(1,2)$. 
One possible explanation is that the stencils used for the 1st and 2nd order Taylor terms have an error of 2nd and 3rd order. This could introduce errors which become larger than the Taylor expansion error (see also Section \ref{subsec:taylor}). We have therefore recomputed the results for $(r,s) = (1,2)$ and $(r,s)=(1,3)$ using stencils that are accurate to 3rd and 4th order, resp. The resulting errors are -0.163\,bp / $-0.033\%$ and -0.061\,bp / $-0.012\%$. The accuracy decreased for $s=2$ and increased slightly for $s=3$. However,  for $r=1$ all approximations for $s=1,2,3$ are now all below the full value and seem to converge towards it with increasing $s$. 

We also note that some of the computed terms for large $r+s$ are smaller than their standard deviation of around 0.002. Although the leading digit may not be significant, the point stands that these terms are small.

A general comment is due on the computation of the values in Tables \ref{tab:RFhigherorder2} and \ref{tab:RFhigherorder}.
For demonstration purposes, instead of solving PDEs as previously, we have written the solutions as expectations (in the obvious way using the Feynman-Kac theorem) and estimated them by
Monte Carlo simulation. 
For each triple of $r$, $s$ and $t$ values we have used the same Brownian paths for all the terms on the right-hand side of (\ref{Vrst}).
This has a two-fold benefit. Firstly, it reduces the variance of the estimator for these terms, by a similar mechanism as for standard finite difference sensitivies
(see \cite{G03}). Note that a large number of these finite differences are to be computed here, in particular $N-r$ choose $s$ terms of highest order, and these normally have mixed signs. Secondly, recycling the normally distributed samples reduces the computation time.

While this gives us a computationally convenient way to illustrate the behaviour of these terms, 
this is not how one would solve this problem in practice, since a direct simulation of the full problem is possible in this case.
The full benefit of the expansion method is realised in cases where accurate MC solutions are not or not easily available, but accurate PDE solutions to low-dimensional approximations are feasible --- such as the Bermudan swaption investigated earlier.
(But since a sufficiently accurate and reliable alternative solution to the full problem is not available in those cases, we have no benchmark to compare against and therefore omitted these computations for the purposes of this study.)

We also anticipate that there are advantages in
the use of hybrid methods. In these, the 
lower order terms can be computed very precisely with PDE methods and the 
higher order terms, which are much smaller in size and often show a corresponding decline in their variance, are computed with reduced relative accuracy via MC simulation. Due to the decreased requirements on the relative accuracy, the latter may use a comparatively low number of MC paths.

Additionally, MC simulation can be used for those higher-dimensional terms in cases where it only provides a crude approximation, such as the lower bound for the Bermudan Swaptions in the previous section. For example, if the PDE based lower order terms are already correct within $0.1\%$ and the higher order MC based terms that provide a correction of that size are only accurate to within $10\%$, their inclusion will still reduce the overall error by an order of magnitude.

\section{Discussion and outlook}\label{sec:Con}


The results presented in this article demonstrate the practical applicability of a systematic expansion approach
to the LIBOR Market Model. 

\subsubsection*{Summary and discussion}

We were able to compute values for Bermudan swaptions and Ratchet floors 
which showed a very good match to the Monte Carlo benchmark for up to $N=40-60$ quarterly LIBORs
under a range of market conditions.
The run times were of the same order of magnitude  as the MC run times for the path-dependent example and substantially smaller for the early exercise case.

In extensions to higher order, we find that both increasing the number $r$ of eigenvalues which are fully included in all expansion terms and increasing the order $s$ leads to a rapid decrease of the relative error, in line with theoretical predictions. The optimal choice of $r$ and $s$ depends on the desired accuracy. Including at least the first order Taylor terms seems generally beneficial, because the corresponding accuracy improvement is significant while keeping the computational cost tractable.

Encouragingly, a closer look at the raw data going into Figure \ref{fig:dimconv} reveals that the correction
terms for decreasing eigenvalues are indeed strictly decreasing. This could be used as the (heuristic) basis for dimension adaptivity.

\subsubsection*{Analysis}

Section \ref{sec:PCAApproach} motivates the PCA-ANOVA approach via Taylor expansion, but further work is needed on the theoretical underpinning of the method. 
In particular, the size of the coefficients in the Taylor and ANOVA expansions depends on the smoothness of the solution and it is ongoing work to derive error bounds. 

In essence, under some technical conditions, piece-wise smoothness of the payoff is sufficient for convergence, with possible problems (or slower convergence) only at the kinks and there only in degenerate cases. For Bermudan and path-dependent cases, some additional complexity arises due to interval/update conditions. However, for typical such conditions, piece-wise smoothness holds and the technical conditions are preserved. We sketch a heuristic analysis here.


Over one period, such as for European-style options,
the solution to (\ref{eq:PDEHeat}) is given by
\begin{eqnarray}
\label{heat-kernel}
u(z,\tau;\lambda) &=& \frac{1}{(2 \pi)^{N/2}}
\int_{\mathbb{R}^N}
\exp\left(-\sum_{i=1}^N \xi_i^2/2 \right) \; u(z + \sqrt{\tau \Lambda} \, \xi,0;\lambda) \; {\rm d}\xi,
\end{eqnarray}
where $\Lambda$ is the diagonal matrix with $\lambda$ in the diagonal.

Consider first the case $r=1$, $s=1$. Virtually all practically relevant payoffs are piecewise smooth with ``kinks'' (i.e., gradient discontinuities), or ``jumps'' (i.e., discontinuities), along curves, surfaces etc.
From the above solution formula (\ref{heat-kernel}) it is clear that the derivatives with respect to $\lambda_i$, $i=2,\ldots,N$ are related to the spatial derivatives of the payoff. (We allow for $\lambda$-dependence of the payoff in preparation for the multi-period case.)
These $\lambda$-derivatives thus exist at $\lambda_2=\ldots=\lambda_N=0$ as long as the diffusion in the direction of the first coordinate (we leave $\lambda_1>0$ fixed and thus the convolution with the heat kernel in direction $\xi_1$ in (\ref{heat-kernel}) remains) provides smoothing, precisely, if the curve, surface etc which describes the location of the kink is not locally parallel to the first coordinate axis. Even if this were to happen, it would only happen for isolated coordinate values. We conjecture that the leading order error term would then not be $\|\lambda-\lambda_0\|^2$, but $\|\lambda-\lambda_0\|$, and that this only appears at isolated spatial coordinates.

The case of larger $r$ and $s$ can be characterised similarly.


For more complex derivatives such as the ones studied in Sections \ref{subsec:BS} and \ref{subsec:ratfloor}, the diffusion equation (\ref{eq:PDEHeat}) holds piecewise in time intervals $(T_i,T_{i+1})$, while at $T_i$ interface conditions hold. The coefficients in the expansion (\ref{taylor}) are now determined by a recursion over $i$. The operations (\ref{ex-cond}) and (\ref{eq:jumpcond}) generate value functions at $T_i$ which are again piecewise smooth -- with kinks -- and serve as terminal conditions for the preceding time interval. The above ideas apply recursively.

\subsubsection*{Variable coefficients}

A practically important extension is to treat variable coefficients in the PDE accurately and systematically. In the model studied here, the covariance matrix was assumed constant and the non-constant drift was approximated by a constant one.
A more general approach to variable coefficients would be to ``freeze'' only a subset of the covariance and drift components as is required for the anchored ANOVA. We expect this
to give higher order accuracy in $T$.
This will allow us to use more complex volatility and correlation structures than the ones described in Section \ref{sec:PCALMM}.

\subsubsection*{Conclusion}

Overall, we believe that the approach discussed here can be developed into an extremely powerful and versatile framework for the approximation of high-dimensional problems. It is not inherently restricted to the LIBOR market or mathematical finance problems in general; we expect it to perform well across a wide range of problems with suitable correlation structures.

\bibliographystyle{plain}
\bibliography{Bibliography}

\appendix

\section{Further results}
\label{app:further}

\begin{table}[htbp]
\begin{center}
\begin{tabular}{cccccccccc}
$N$ & $V_{MC}^-$ & $\sigma$ & $V_{MC}^- + \Delta_0$ & $\sigma_{\Delta_0}$ & $V_{MC}^- + \Delta_0/2$ & $V_{PDE}$ & $\Delta_{abs}$ & $\Delta_{rel}$ \\ \hline\hline
5 & 1.16E-02 & 2.36E-12 & 1.16E-02 & 0.00E+00 & 1.16E-02 & 1.16E-02 & -8.73E-15 & 0.00\% \\ 
11 & 2.38E-02 & 1.29E-12 & 2.38E-02 & 4.74E-06 & 2.38E-02 & 2.38E-02 & -1.18E-05 & -0.05\% \\ 
21 & 4.59E-02 & 1.34E-05 & 4.68E-02 & 4.89E-05 & 4.63E-02 & 4.67E-02 & 4.15E-04 & 0.90\% \\ 
41 & 8.59E-02 & 2.13E-05 & 8.93E-02 & 1.21E-04 & 8.76E-02 & 8.93E-02 & 1.70E-03 & 1.94\% \\ 
\hline
5 & 1.16E-02 & 2.36E-12 & 1.16E-02 & 0.00E+00 & 1.16E-02 & 1.16E-02 & -8.73E-15 & 0.00\% \\ 
11 & 2.38E-02 & 1.29E-12 & 2.38E-02 & 4.80E-06 & 2.38E-02 & 2.38E-02 & -1.37E-05 & -0.06\% \\ 
21 & 4.59E-02 & 1.34E-05 & 4.68E-02 & 5.08E-05 & 4.64E-02 & 4.67E-02 & 3.72E-04 & 0.80\% \\ 
41 & 8.59E-02 & 2.15E-05 & 8.95E-02 & 1.29E-04 & 8.77E-02 & 8.93E-02 & 1.64E-03 & 1.87\% \\ 
 \end{tabular}
\end{center}
\caption{PDE results for ITM ($K=0.09$) Bermudan swaptions compared to frozen (top) and full (bottom) drift MC results.}
\label{tab:ITM}
\end{table}

\begin{table}[htbp]
\begin{center}
\begin{tabular}{cccccccccc}
$N$ & $V_{MC}^-$ & $\sigma$ & $V_{MC}^- + \Delta_0$ & $\sigma_{\Delta_0}$ & $V_{MC}^- + \Delta_0/2$ & $V_{PDE}$ & $\Delta_{abs}$ & $\Delta_{rel}$ \\ \hline\hline
5 & 9.39E-04 & 6.78E-07 & 9.39E-04 & 0.00E+00 & 9.39E-04 & 9.49E-04 & 9.97E-06 & 1.06\% \\ 
11 & 7.01E-03 & 4.15E-06 & 7.08E-03 & 6.51E-06 & 7.04E-03 & 7.30E-03 & 2.57E-04 & 3.64\% \\ 
21 & 2.02E-02 & 9.60E-06 & 2.17E-02 & 6.29E-05 & 2.10E-02 & 2.11E-02 & 1.01E-04 & 0.48\% \\ 
41 & 4.50E-02 & 1.74E-05 & 5.12E-02 & 1.77E-04 & 4.81E-02 & 4.91E-02 & 9.90E-04 & 2.06\% \\ 
\hline
5 & 9.39E-04 & 6.78E-07 & 9.39E-04 & 0.00E+00 & 9.39E-04 & 9.49E-04 & 1.02E-05 & 1.09\% \\ 
11 & 7.03E-03 & 4.16E-06 & 7.11E-03 & 6.70E-06 & 7.07E-03 & 7.30E-03 & 2.30E-04 & 3.25\% \\ 
21 & 2.03E-02 & 9.66E-06 & 2.19E-02 & 6.31E-05 & 2.11E-02 & 2.11E-02 & -4.85E-05 & -0.23\% \\ 
41 & 4.55E-02 & 1.77E-05 & 5.22E-02 & 1.85E-04 & 4.88E-02 & 4.91E-02 & 2.62E-04 & 0.54\% \\ 
\end{tabular}
\end{center}
\caption{PDE results for OTM ($K=0.11$) Bermudan swaptions compared to frozen (top) and full (bottom) drift MC results.}
\label{tab:OTM}
\end{table}

\begin{table}[htbp]
\begin{center}
\begin{tabular}{ccccccc}
$(a,b,c)$ & $N$ & $V_{MC}$ & $\sigma$ & $V_{PDE}$ & $\Delta_{abs}$ & $\Delta_{rel}$ \\ \hline\hline
$K_1 = 0.10$ &&&&&&\\
0/1/0 & 5 & 7.08E-003 & 2.95E-006 & 7.04E-003 & -3.25E-05 & -0.46\% \\ 
 & 11 & 9.63E-003 & 3.86E-006 & 9.61E-003 & -2.18E-05 & -0.23\% \\ 
 & 21 & 1.06E-002 & 4.17E-006 & 1.07E-002 & 9.78E-05 & 0.92\% \\ 
0.2/0.9/0 & 5 & 3.06E-002 & 4.23E-006 & 3.06E-002 & -6.80E-05 & -0.22\% \\ 
 & 11 & 4.94E-002 & 4.73E-006 & 4.93E-002 & -1.04E-04 & -0.21\% \\ 
 & 21 & 5.10E-002 & 5.02E-006 & 5.09E-002 & -1.06E-04 & -0.21\% \\ 
0.25/0.95/-0.01 & 5 & 3.29E-002 & 4.03E-006 & 3.29E-002 & -6.26E-05 & -0.19\% \\ 
 & 11 & 6.06E-002 & 5.13E-006 & 6.06E-002 & -2.65E-05 & -0.04\% \\ 
 & 21 & 7.37E-002 & 8.14E-006 & 7.36E-002 & -1.01E-04 & -0.14\% \\ 
\hline
$K_1 = 0.11$ &&&&&&\\
	0/1/0 & 5 & 1.27E-002 & 3.88E-006 & 1.26E-002 & -5.16E-05 & -0.41\% \\ 
	 & 11 & 1.44E-002 & 4.73E-006 & 1.43E-002 & -6.55E-05 & -0.45\% \\ 
	 & 21 & 1.44E-002 & 4.91E-006 & 1.44E-002 & 1.01E-05 & 0.07\% \\ 
	0.2/0.9/0 & 5 & 3.63E-002 & 4.33E-006 & 3.63E-002 & -6.70E-05 & -0.18\% \\ 
	 & 11 & 5.21E-002 & 4.76E-006 & 5.20E-002 & -1.04E-04 & -0.20\% \\ 
	 & 21 & 5.18E-002 & 5.02E-006 & 5.17E-002 & -9.75E-05 & -0.19\% \\ 
	0.25/0.95/-0.01 & 5 & 4.00E-002 & 4.11E-006 & 4.00E-002 & -6.16E-05 & -0.15\% \\ 
	 & 11 & 6.52E-002 & 5.12E-006 & 6.52E-002 & -2.53E-05 & -0.04\% \\ 
	 & 21 & 7.58E-002 & 8.09E-006 & 7.57E-002 & -1.03E-04 & -0.14\% \\ 
\hline
$K_1 = 0.09$ &&&&&&\\
	0/1/0 & 5 & 3.19E-003 & 1.94E-006 & 3.17E-003 & -1.81E-05 & -0.57\% \\ 
	 & 11 & 5.82E-003 & 2.94E-006 & 5.82E-003 & -2.70E-06 & -0.05\% \\ 
	 & 21 & 7.33E-003 & 3.40E-006 & 7.47E-003 & 1.41E-04 & 1.92\% \\ 
	0.2/0.9/0 & 5 & 2.51E-002 & 4.07E-006 & 2.50E-002 & -5.99E-05 & -0.24\% \\ 
 	& 11 & 4.68E-002 & 4.70E-006 & 4.67E-002 & -1.14E-04 & -0.24\% \\ 
	 & 21 & 5.03E-002 & 5.02E-006 & 5.02E-002 & -1.00E-04 & -0.20\% \\ 
	0.25/0.95/-0.01 & 5 & 2.59E-002 & 3.88E-006 & 2.58E-002 & -5.57E-05 & -0.22\% \\ 
	 & 11 & 5.61E-002 & 5.14E-006 & 5.60E-002 & -3.09E-05 & -0.06\% \\ 
	 & 21 & 7.15E-002 & 8.20E-006 & 7.15E-002 & -6.08E-05 & -0.09\% \\ 
\end{tabular}
\end{center}
\caption{PDE results for Ratchet floors compared to frozen drift MC results. The column $V_{PDE}$ shows the computed PDE value. Columns $\Delta_{abs}$ and $\Delta_{rel}$ show the absolute and relative difference to the MC estimate $V_{MC}$, resp.}
\label{tab:RF}
\end{table}

\end{document}